\documentclass[letterpaper, 10 pt, conference]{ieeeconf}

\usepackage{cite}
\usepackage{amsmath,amssymb,amsfonts,amsthm}
\usepackage{graphicx}
\usepackage{mathtools}
\usepackage{booktabs}

\theoremstyle{plain}
\newtheorem{prop}{Proposition}
\newtheorem{lem}{Lemma}

\theoremstyle{remark}
\newtheorem{rem}{Remark}

\IEEEoverridecommandlockouts 

\begin{document}
	\title{\Huge Nonlinear modeling and feedback control of \\ boom barrier automation}
	
	\author{Daniel~Cunico,
		Angelo~Cenedese,
		Luca~Zaccarian
		and~Mauro~Borgo
		\thanks{D. Cunico, A. Cenedese are with the Department of Information Engineering, University of Padua, Padova, Italy, 35131, Email: (daniel.cunico@studenti.unipd.it; angelo.cenedese@unipd.it).}
		\thanks{D. Cunico, M. Borgo are with BFT SpA, Schio (VI), Italy, 36015, Email: (daniel.cunico@bft-automation.com; mauro.borgo@bft-automation.com).}
		\thanks{L. Zaccarian is with the Department of Industrial Engineering, University of Trento, Italy, and LAAS-CNRS, Université de Toulouse, CNRS, Toulouse, France, Email: zaccarian@laas.fr}
	}

	\maketitle

\begin{abstract}
	We address modeling and control of a gate access automation system. A model of the mechatronic system is derived and identified. Then an approximate explicit feedback linearization scheme is proposed, which ensures almost linear response between the external input and the delivered torque.
	 A nonlinear optimization problem is solved offline to generate a feasible trajectory associated with a feedforward action and a low level feedback controller is designed to track it. 
	 The feedback gains can be conveniently tuned by solving a set of convex linear matrix inequalities, performing a multi-objective trade-off between disturbances attenuation and closed-loop performance. Finally, the proposed control strategy is tested on the real system and experimental results show that it can effectively meet the requirements in terms of robustness, load disturbance rejection and tracking performance.
\end{abstract}

\section{Introduction}

Due to the increasing global industrial competition for extreme performance and reliability,
research on mechatronic systems is becoming highly multidisciplinary, with an ever-increasing integration of mechanical, electronic, and information disciplines \cite{Isermann}. Nonlinear actuator phenomena, such as saturations, dead-zones, backlash or sampling/quantization effects, appear frequently in mechatronic systems \cite{Nordin, Zaccarian, Corradini} and have proved to be a source of performance degradation and closed-loop instability. The efficiency highly depends on the control architecture and its ability to consider the limitations and constraints in order to optimize the functioning and to avoid dangerous working conditions. Moreover, robust control design techniques \cite{lin2007robust} are key for
taking into account uncertain model parameters within a broad range of real situations, e.g. due to changes in the environmental conditions or wear of the mechanical components.  

Access automation systems are used in several residential or public areas to prevent unwanted access or to regulate traffic flow \cite{hotto2007motion, jones2008gate}. An active industrial research area deals with the performance and quality improvement for such systems, while lowering the manufacturing costs and the power consumption. 

The standard control techniques are those typical of electromechanical motion systems \cite{levine, LAMBRECHTS, 491410, 1104009} comprising two hierarchical levels: a {\em trajectory planner} generates the desired reference, taking into account the nonlinear dynamics and the constraints; a {\em linear error feedback} 
reduces the deviation of the actual trajectory from the desired reference. 
The current industrial practice for parameter tuning is based on running several experimental tests and adjusting certain PID gains via trial and error, until acceptable results are obtained, thus requiring much time and man power from the technical department.

\begin{figure}[t!]
\centering 
\includegraphics[width=0.45\textwidth]{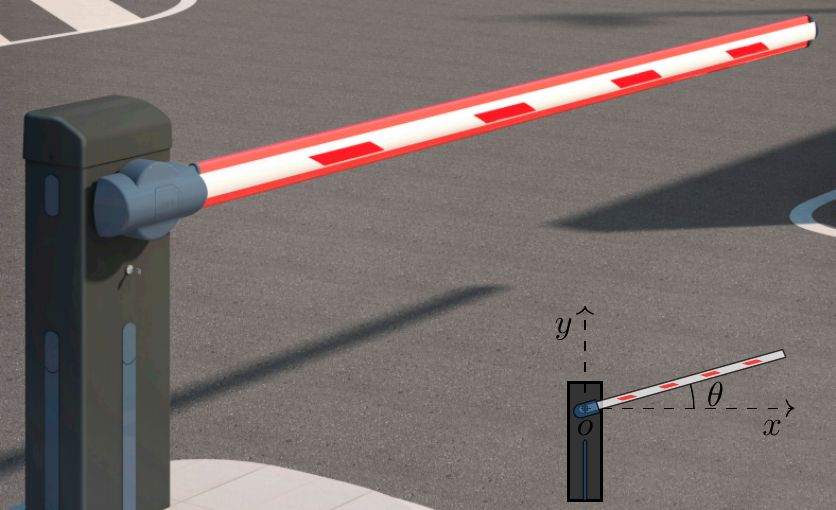}
\caption{The boom barrier experimental system.}
\label{barrier_fig}
\end{figure}

In this paper, we focus on the modeling and control of a road automatic barrier represented in Fig. \ref{barrier_fig}.  
The original contributions of this work are highlighted next. 1) First we 
derive and experimentally validate a nonlinear mathematical model of the underlying unidirectional power converter, the electrical motor and the mechanical transmission moving the load by well representing the 
interplay of mechanical and electrical components (the mechatronic device).
2) Secondly, we propose an approximate feedback inversion scheme, whose effectiveness is proven by relying on formally certified interval arithmetic combined with formal Taylor expansion (thanks to the Coq Interval tactic \cite{martin2016proving}): through this scheme, we include a feedback linearizing pre-compensator, precisely characterizing the state-dependent saturation values of the virtual input proportional to the exerted voltage. 3) Thirdly, based on this feedback linearizing structure we propose a feedforward/feedback architecture, whose feedforward term is generated
through the minimization of a nonlinear functional cost under constraints,
and the feedback term is conveniently tuned via a linear matrix inequality (LMI) formulation \cite{Boyd}. The LMI constraints allow us to optimize a disturbance rejection performance 
under uncertain model parameters, while constraining the closed-loop poles in a suitable region of the left half-plane \cite{LMI2} to induce a suitable transient response.
4) Fourth, rigorous statements certify the effectiveness of our scheme in terms of stabilization of the error dynamics and feasibility of our LMIs.
5) Lastly, and most importantly, experimental results on the industrial device confirm the effectiveness  of the proposed strategy, which induces regular (no oscillations) and fast barrier opening, despite the system nonlinear dynamics and uncertain parameters. 
Beyond the performance improvement in the specifically considered application, the approach is of general interest and it can be easily extended to many similar applications that use the same control electronics.
Some technological details are omitted and all the units of measure are normalized in the experimental results for reasons of confidentiality. However, the proposed design strategy is fully parametric and has been tested successfully with many different parameter selections.
The paper is organized as follows. In Section II the experimental setup is described and the closed-loop goals are clarified. In Section III the mathematical modeling of the road barrier gate is derived, considering all the mechatronic components. In Section IV the augmented plant model and the parameter identification procedure are illustrated. Section V describes the electrical drive and the feedback linearization method. In Section VI the control architecture is presented and an LMI based tuning procedure is proposed. Experimental tests are discussed in Section VII. Concluding remarks are reported in Section VIII.

\section{System Description and Goals}
\begin{figure}
	\centering
	\includegraphics[width=0.45\textwidth]{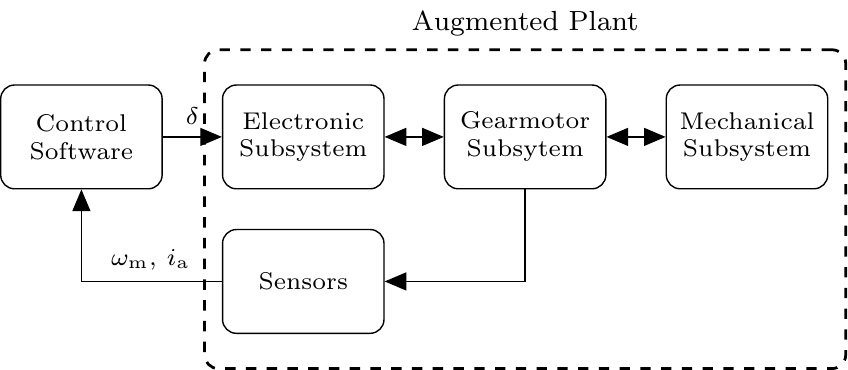}
	\caption{Blocks diagram of a common mechatronic system.}
	\label{mechatronic_system_fig}
\end{figure}

The considered mechatronic system can be represented as sketched in Fig. \ref{mechatronic_system_fig}. The electronic parts are the power source circuit and the driver of the motor. A DC motor converts electrical energy into mechanical energy and produces the torque required to move the load with the desired output angular speed. The torque is transmitted through a gearbox to the mechanical system. Two main elements compose the mechanics of the automatic road barrier: a bar that rotates about one of its ends, and a spring-damper system used to compensate for the weight of the bar. The sensor devices represent the part related to the data acquisition system, i.e. the group of sensors and transducers with their conditioning circuits. In the present case study, the acquired measurements are the motor speed $\omega_\mathrm{m}$ and the motor current $i_\mathrm{a}$. The electronic, gearmotor and mechanical subsystems together with the sensors form the so-called augmented plant. Finally, the embedded control software produces the duty cycle $\delta$ of a PWM signal to control the actuator with precise timing. 

The main problems and goals regarding the control of this application can be summarized in the following points:
\begin{enumerate}
	\item (Limitations) The low-cost electronic board does not allow exerting a motor torque/current in the braking direction. Therefore, the braking phase is often slow and the system only decelerates due to the action of friction.
	\item (Safety) The gate opening maneuver must end with a sufficient low speed at the mechanical stop in order to avoid damaging the device.
	\item (Performance) The gate opening should be regular (without oscillations) and fast.
\end{enumerate}
In addition to the aims defined above, the controller should be robust with respect to possible slow unmodeled dynamics, small delays in the loop, quantization effects, variations related to environmental conditions and aging. Finally, the proposed control strategy must be easy enough to be implemented in the micro-controller unit of the industrial device, which has limited computational capacity.

\section{Modeling}

\subsection{Electric motor}
A brushed DC electric motor exerts the torque on the mechanical subsystem. The DC motor dynamic model is well known in the literature \cite{krishnan}. The electrical equation is:
\begin{equation}
	u_\mathrm{a}(t) = R_\mathrm{a} i_\mathrm{a}(t) +L_\mathrm{a} \frac{di_\mathrm{a}(t)}{dt} + e_\mathrm{a}(t),
	\label{elec_eq_motor}
\end{equation}
where $u_\mathrm{a}(t)$ is the terminal voltage, $i_\mathrm{a}(t)$ is the armature current, $R_\mathrm{a}$ is the armature winding resistance, $L_\mathrm{a}$ is the phase inductance and $e_\mathrm{a}(t)$ is the back electromotive force (BEMF). 
The BEMF and the torque exerted at the motor shaft correspond to
\begin{align}
	e_\mathrm{a}(t) &= k_\mathrm{t} \omega_\mathrm{m}(t), \label{BEMF_eq} \\
	\tau_\mathrm{m}(t) &=k_\mathrm{t} i_\mathrm{a}(t),	\label{torque_eq}	
\end{align}
where $k_\mathrm{t}$ is the torque constant and $\omega_\mathrm{m}$ is the mechanical speed of the motor. Note that the two constants in eq. (\ref{BEMF_eq}) and eq. (\ref{torque_eq}) coincide because of the balance between the input electrical power and the output mechanical power.

\subsection{Mechanical system}
\begin{figure}[t!]
	\centering
	\includegraphics[width=0.45\textwidth]{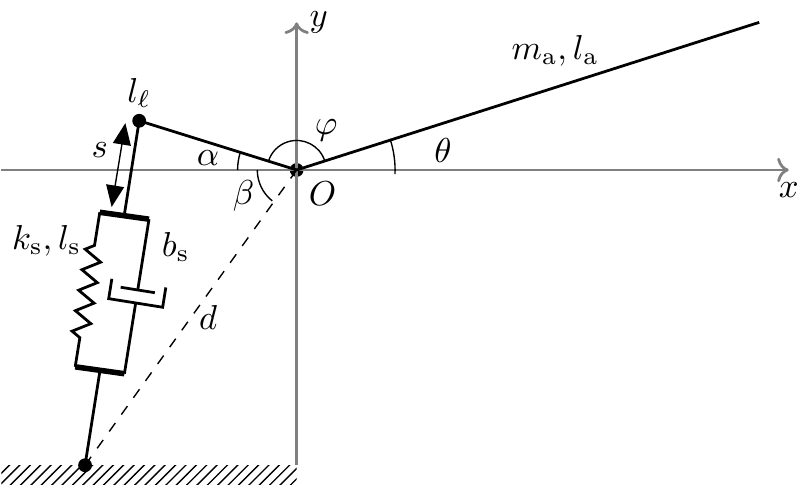}
	\caption{Mechanical subsystem of the automatic barrier.}
	\label{mechanical_system_fig}	
\end{figure}

The mechanical subsystem of the automatic road barrier, represented in Fig. \ref{mechanical_system_fig}, is composed by two main elements:\\
1)	a bar rotating about one of its ends (the point $O$), assumed to be an ideal rod of length $l_\mathrm{a}$ and mass $m_\mathrm{a}$, whose angular position with respect to the $x$-axis in Fig. \ref{mechanical_system_fig} is described by the angle $\theta$.\\
2) a spring-damper of natural length $l_{\rm s,0}$ and spring constant $k_\mathrm{s}$, with one of its ends connected to the bar through a lever of length $l_\mathrm{\ell}$. The lever element is fixed to the bar in $O$, thus forming with it a constant angle $\varphi$. The damper element produces a force proportional to the velocity, according to the viscous coefficient $b_\mathrm{s}$, allowing for the stabilization of the entire mechanical system. 

Furthermore, it is possible to pre-compress the spring of a length $s_\mathrm{0}$ in order to calibrate the resulting force. Typically $s_\mathrm{0}$ is tuned in such a way that the entire system be at the equilibrium when $\theta = \theta_{\rm e}=\pi/4$. We assume that the mass of the spring and of the lever are negligible. From geometric considerations, we obtain the following expressions for the angle $\alpha(\theta)$ of the lever w.r.t. the $x$-axis, the length $l_\mathrm{s}(\theta)$ of the spring and the compression $s(\theta)$ of the spring:
\begin{align*}
	\alpha(\theta) &= \pi - \varphi -\theta, \\
	l_\mathrm{s}(\theta) &= \sqrt{d^2 +l_\mathrm{\ell}^2 - 2 d \, l_\mathrm{\ell}\text{cos}\left( \beta + \alpha(\theta) \right) }, \\
	s(\theta) &=  l_{\rm s,0} - l_\mathrm{s}(\theta) +s_\mathrm{0}.
\end{align*}
Following the notation used in Fig. \ref{mechanical_system_fig}, the inertia and the friction of the mechanical load are
\begin{subequations}
\begin{equation}
	J_\mathrm{a} = \frac{1}{3}m_\mathrm{a} l_\mathrm{a}^2, \quad b(\theta) = b_\mathrm{s} \frac{l_\mathrm{\ell} d}{l_\mathrm{s}(\theta)} \text{sin} \left(\beta + \alpha(\theta) \right),
	\label{mech_system_eq}
\end{equation}
and the reaction torque exerted by the rod at the hinge corresponds to
\begin{equation}
	\tau_\mathrm{r}(\theta) = \underbrace{- k_\mathrm{s} s(\theta)  \frac{l_\mathrm{\ell} d}{l_\mathrm{s}(\theta)} \text{sin} \left(\beta + \alpha(\theta) \right)}_{\text{spring torque}} + \underbrace{\frac{g}{2} m_\mathrm{a} l_\mathrm{a} \text{cos}(\theta)}_{\text{bar torque}}.
	\label{eq_tau_r}
\end{equation}
\end{subequations}

\begin{figure}[t]
	\centering
	\includegraphics[width=0.45\textwidth]{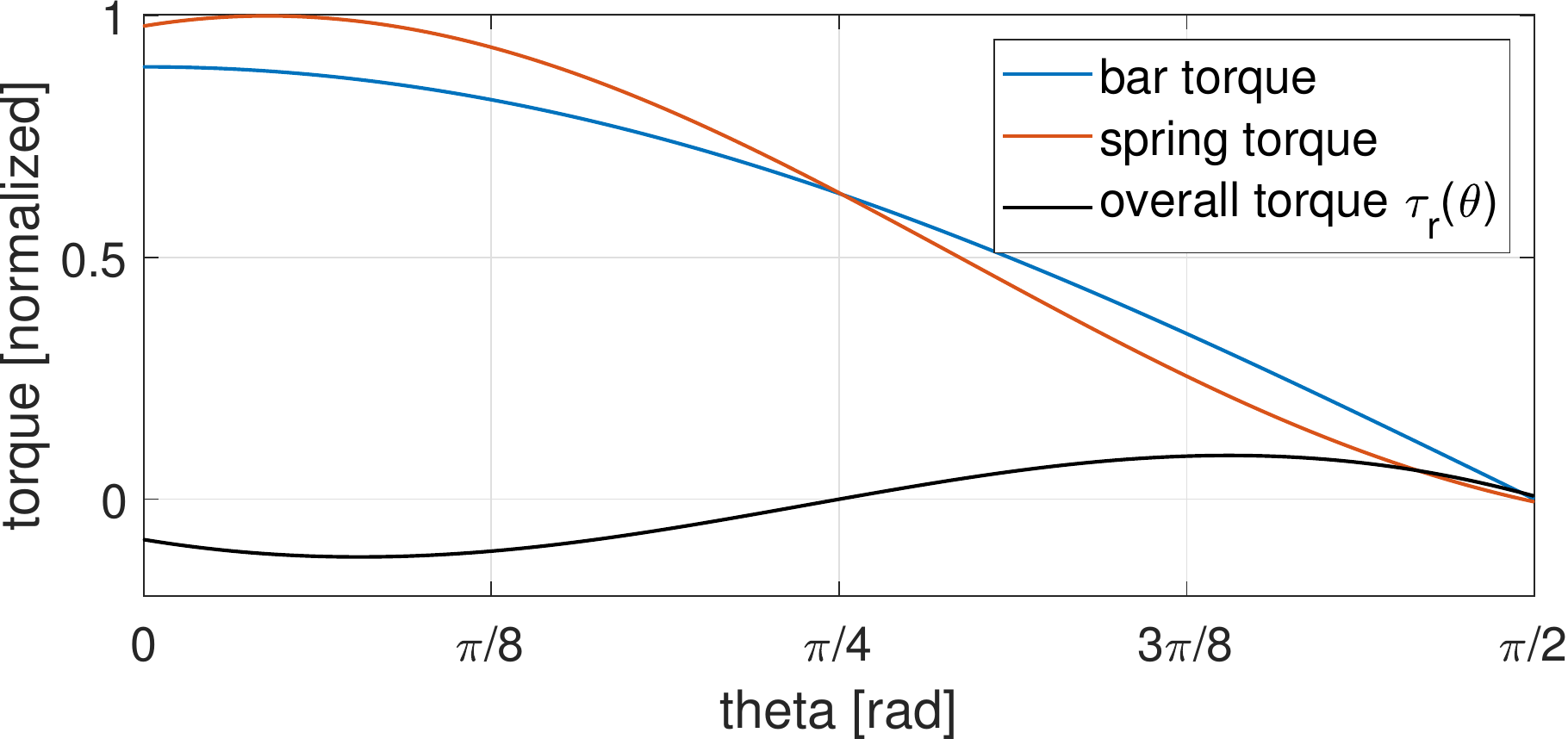}
	\caption{Main components of the torque $\tau_\mathrm{r}(\theta)$ in (\ref{eq_tau_r}). The spring is pre-compressed of a length $s_0$ so that $\theta_{\rm e}=\pi/4$.}
	\label{tau_r_fig}
\end{figure}

\noindent
Fig. \ref{tau_r_fig} shows the evolution of the overall external torque $\tau_\mathrm{r}$ as a function of the angle $\theta$. The first and the second term at the right-hand side of equation (\ref{eq_tau_r}) are respectively the spring and bar contributions to the torque, tuned to generate an equilibrium point at $\theta = \theta_{\rm e}$.

\subsection{Mechanical transmission}

The mechanical transmission consists of a gear train system. The gearbox is modelled by means of the classical mechanical approach assuming rigid coupling \cite{leonhard2001}. In an ideal transmission, i.e. under the assumption of lossless power transfer, denoting by $\omega(t)=\dot{\theta}(t)$ the speed at the output of the gear, we have that:
\begin{equation}
	\omega = \frac{r_1}{r_2} \omega_\mathrm{m} = N_\mathrm{g} \omega_\mathrm{m}, \quad \theta = N_\mathrm{g} \theta_\mathrm{m}
	\label{gear_ratio}
\end{equation}
where $r_1$ and $r_2$ are the gear wheels radii and $N_\mathrm{g}$ is the transmission gear ratio. A better description is achieved by considering an efficiency $\eta<1$ of the transmission gear, and characterizing load torque $\tau_\ell$ as
\begin{equation}
	\tau_\mathrm{\ell}(\theta_\mathrm{m}(t),\omega_\mathrm{m}(t)) = \tau_\mathrm{r}(N_\mathrm{g} \theta_\mathrm{m}(t)) \frac{N_\mathrm{g}}{\eta} + \tau_\mathrm{c} \text{sign}(\omega_\mathrm{m}(t)),
	\label{tau_l}
\end{equation}
where $\tau_\mathrm{r}$ is defined in eq. (\ref{eq_tau_r}) and $\tau_\mathrm{c}$ represents the Coulomb friction torque \cite{makkar2005}. The resulting mechanical equation of the system is
\begin{align}
	\tau_\mathrm{m}(t) = J_{\rm tot} \frac{d\omega_\mathrm{m}(t)}{dt} + b_{\rm tot}\omega_\mathrm{m}(t) + \tau_\mathrm{\ell}(\theta_\mathrm{m}(t),\omega_\mathrm{m}(t))
	\label{gearmotor_eq} \\
	J_{\rm tot} = J_{\rm mg}+ J_\mathrm{a} \frac{N_\mathrm{g}^2}{\eta},  \qquad 
	b_{\rm tot} =  b_{\rm mg} + b(\theta_\mathrm{m}) \frac{N_\mathrm{g}^2}{\eta}.
	\label{mech_eq}
\end{align}
where $J_{\rm mg}$ and $b_{\rm mg}$ are the gearmotor inertia and friction.

\section{Augmented Plant and Identification}
\label{Identification}
\begin{figure*}[t]
	\centering 
	\includegraphics[width=0.42\textwidth]{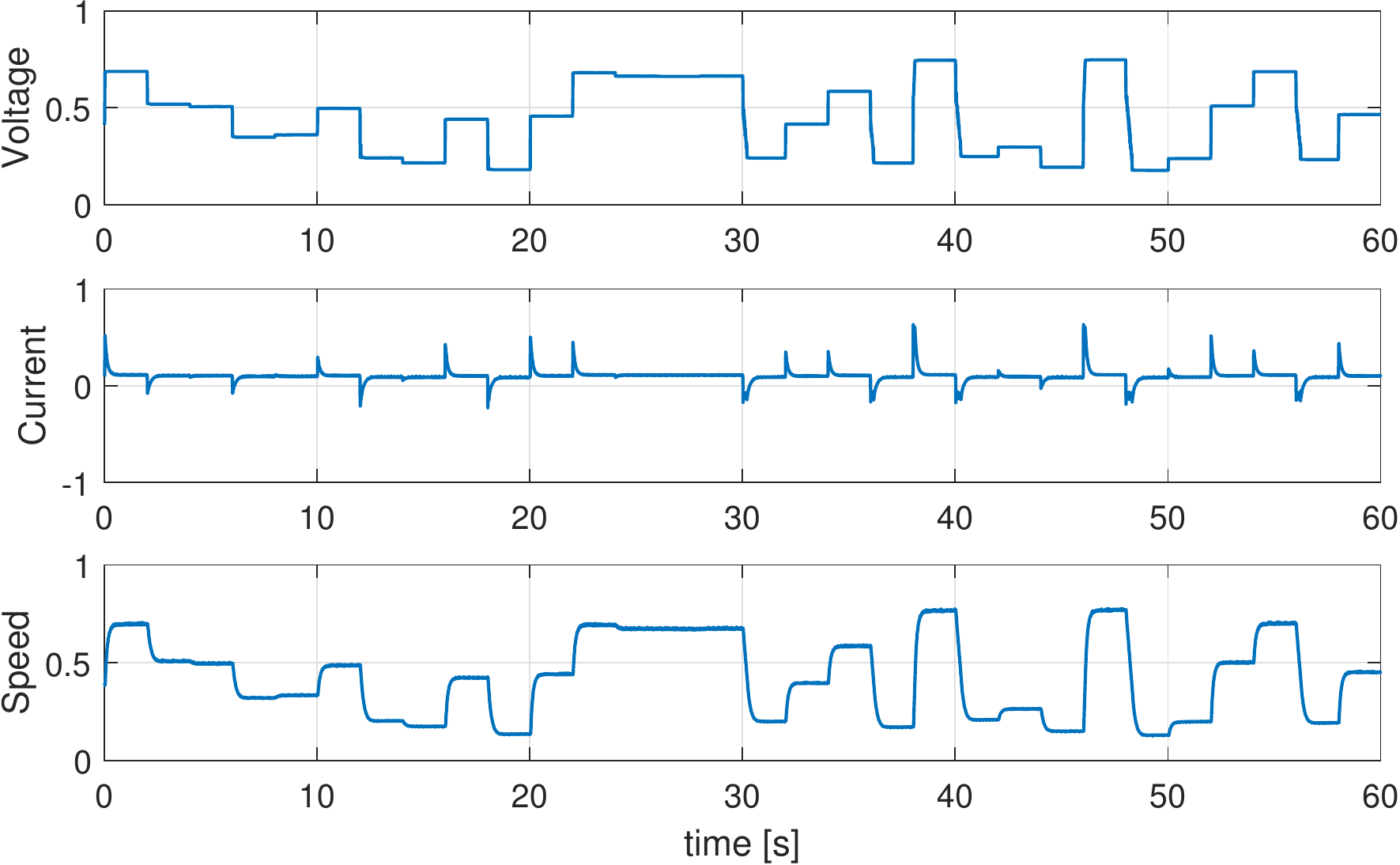}\label{model_id} \qquad
	\includegraphics[width=0.42\textwidth]{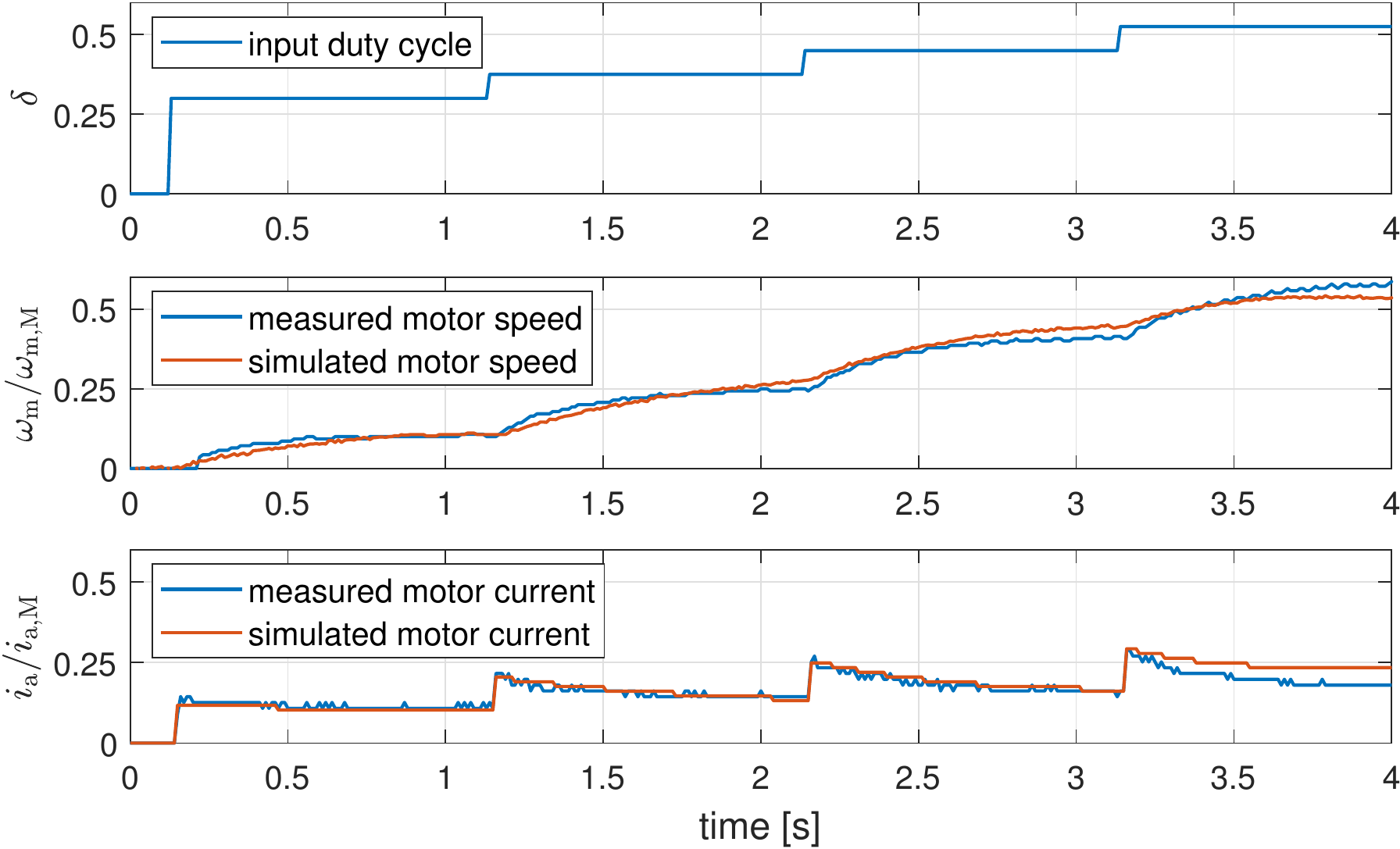}\label{model_val}
	\caption{(Left) Example of data acquired for the identification of the gear-motor parameters: from above to below, respectively, voltage supplied to the motor, motor current and motor speed. (Right) Experimental response to a staircase input duty cycle $\delta$, compared to the identified model simulation. The output signals are the motor speed $\omega_\mathrm{m}$ and the motor current $i_\mathrm{a}$, while $\omega_{\rm{m,M}}$ and $i_{\rm{a,M}}$ are normalization factors.}
	\label{system_id}
\end{figure*}

\subsection{Augmented Plant modeling}
Combining (\ref{elec_eq_motor}), (\ref{BEMF_eq}), (\ref{torque_eq}), (\ref{gearmotor_eq}) and recalling that $\theta=N_\mathrm{g} \theta_\mathrm{m}$, a state-space model of the augmented plant can be obtained. Denoting by $u_\mathrm{a}$ the voltage applied to the motor terminals, by $x_1=i_\mathrm{a}$ the motor current, by $x_2 = \theta_\mathrm{m}$ the motor position and by $x_3 = \omega_\mathrm{m}$ the motor velocity, we have, with $x=\left[x_1, \, x_2, \, x_3\right]^\top$,
\begin{equation}
	\dot{x}=f(x,u_\mathrm{a}) = \begin{dcases}
		& - \frac{R_\mathrm{a}}{L_\mathrm{a}}x_1 - \frac{k_\mathrm{t}}{L_\mathrm{a}}x_3 + \frac{1}{L_\mathrm{a}} u_\mathrm{a} \\
		& x_3 \\
		& \frac{k_\mathrm{t}}{J_{\rm tot}}x_1 - \frac{b_{\rm tot}(x_2)}{J_{\rm tot}}x_3 - \frac{\tau_\mathrm{\ell}(x_2,x_3)}{J_{\rm tot}}
	\end{dcases}
	\label{augmented_plant_model}
\end{equation}
where $\tau_\mathrm{\ell}(x_2,x_3) = \tau_\mathrm{r}(N_\mathrm{g} x_2) \frac{N_\mathrm{g}}{\eta} + \tau_\mathrm{c} \text{sign}(x_3)$ according to \eqref{tau_l}.
Table \ref{table_parameters} reports all the relevant quantities appearing in (\ref{mech_eq}), (\ref{augmented_plant_model}), and their definitions.
\begin{table}
	\caption{Parameters of model (\ref{mech_eq}), (\ref{augmented_plant_model}).}
	\label{table_parameters}
	\centering
	\begin{tabular}{lll}
		\toprule
		Symbol & Name & Defined in  \\ 
		\midrule 
		$R_\mathrm{a}$ & Armature resistance [$\Omega$] & eq. (\ref{elec_eq_motor}) \\
		$L_\mathrm{a}$ & Armature inductance [$H$] & eq. (\ref{elec_eq_motor}) \\ 
		$k_\mathrm{t}$ & Torque constant [$Nm/A$] &  eq. (\ref{torque_eq}) \\ 
		$N_\mathrm{g}$ & Gear ratio & eq. (\ref{gear_ratio}) \\ 
		$\eta$ & Gear efficiency & eq. (\ref{tau_l}) \\
		$b_{\rm mg}$ & Gearmotor viscous friction [$Nms$] & eq. (\ref{gearmotor_eq}) \\ 
		$J_{\rm mg}$ & Gearmotor inertia [$Kgm^2$] & eq. (\ref{gearmotor_eq}) \\
		$J_\mathrm{a}$ & Rod inertia [$Kgm^2$] & eq. (\ref{mech_system_eq}) \\
		$b$ & Nonlinear spring damping [$Nms$] & eq. (\ref{mech_system_eq}) \\ 
		$\tau_\mathrm{r}$ & Reaction torque [$Nm$] & eq. (\ref{eq_tau_r}) \\
		$\tau_\mathrm{c}$ & Coulomb friction torque [$Nm$] & eq. (\ref{augmented_plant_model}) \\ 
		\bottomrule
	\end{tabular}
\end{table}

\subsection{System identification}
\label{section:model_identification}
The focus is now on the identification of the model parameters and their experimental validation. To this aim, well-established System Identification methods \cite{ljung} are used. The nominal model parameters are provided from the literature and the technical data-sheet, while other parameter values can be estimated from the experimental data, to accurately describe the system response. Following \cite{beghi1}, for the electrical and mechanical parameters of the motor and gear subsystem, a set of experiments has been performed with an independent laboratory acquisition system to measure the responses to canonical input signals. For identification purposes, the sampling frequency is 20 times higher than that of the industrial product and a 16-bit ADC resolution is used. A National Instruments DAQ board (USB-6216) has been used to acquire the data and the motor has been equipped with a 12-bit resolution encoder (Eltra ER38F).
Voltage, current and speed are continuous-time signals acquired through a sampling that produces two discrete-time datasets of length $n$.

For the estimation of the electrical parameters $R_\mathrm{a}$, $L_\mathrm{a}$ of eq. (\ref{elec_eq_motor}), the following equation has been identified:
\begin{equation}
	\frac{di_\mathrm{a}(t)}{dt} = -\frac{R_\mathrm{a}}{L_\mathrm{a}} i_\mathrm{a}(t) +  \frac{1}{L_\mathrm{a}} u_\mathrm{a}(t), \:\: y_\mathrm{a}(t) = i_\mathrm{a}(t-\Delta)
	\label{iden_1}
\end{equation}
where $\Delta$ is the delay due to the laboratory acquisition system. 
Note that, as compared with (\ref{elec_eq_motor}), $e_\mathrm{a}$ is not present in \eqref{iden_1} since the identification phase is performed under locked rotor condition, that is, from \eqref{BEMF_eq}, $e_\mathrm{a}$ is zero.
Notably, the delay should not be identified since the industrial device does not contain the acquisition system used to perform the parameter identification experiments.
Since input $u_\mathrm{a}$ is constant during each sampling period, the sampled measurements $y_\mathrm{a}(k)$ collected from (\ref{iden_1}) depend on two values of the discretized input as follows
\begin{equation}
	y_\mathrm{a}(k+1) = \Phi y_\mathrm{a}(k) + \Gamma_\mathrm{0} u_\mathrm{a}(k+1) + \Gamma_1 u_\mathrm{a}(k),
	\label{arx_model}
\end{equation}
\begin{align*}
    \Phi &=  e^{-\frac{R_\mathrm{a}}{L_\mathrm{a}} t_\mathrm{s}}, \\
	\Gamma_0 &= \frac{1}{R_\mathrm{a}} \left(1-e^{-\frac{R_\mathrm{a}}{L_\mathrm{a}} (t_\mathrm{s} - \Delta)} \right), \\
	\Gamma_1 &= \frac{1}{R_\mathrm{a}} \left(e^{-\frac{R_\mathrm{a}}{L_\mathrm{a}} (t_\mathrm{s} - \Delta)} - e^{-\frac{R_\mathrm{a}}{L_\mathrm{a}} t_\mathrm{s}}   \right),
\end{align*}
where $u_\mathrm{a}(k)$ is the voltage supplied to the motor at the $k$-th sampling time, $t_k = k t_\mathrm{s}$, $k=1,2,\dots,n$, and $t_\mathrm{s}$ is the sampling time. From eq. (\ref{arx_model}), with a least-squares estimate of the parameters of the model, we can obtain the value of the electrical parameters as:
\begin{equation*}
	R_\mathrm{a}=\frac{1-\Phi}{\Gamma_0 + \Gamma_1}, \quad
	L_\mathrm{a}=-\frac{R_\mathrm{a} \, t_\mathrm{s}}{\text{ln}\Phi}.   
\end{equation*}
In a similar way, for the estimation of the mechanical parameters $b_\mathrm{mg}$, $J_\mathrm{mg}$ of eq. (\ref{mech_eq}), we consider the discretized dynamics between the applied voltage $u_\mathrm{a}$ and the delayed speed measurement $\omega_\mathrm{m}(t-\Delta)$, giving the following equation
\begin{equation}
	\omega_\mathrm{m}(k+1) - \Phi \omega_\mathrm{m}(k) = \Gamma_0 u_\mathrm{a}(k+1) + \Gamma_1 u_\mathrm{a}(k), \label{arx_model2}
\end{equation}
\begin{align*}
	\Phi &=  e^{-\frac{b_\mathrm{mg}R_\mathrm{a}+k_\mathrm{t}^2}{J_\mathrm{mg}R_\mathrm{a}} t_\mathrm{s}}, \\
	\Gamma_0 &= \frac{k_\mathrm{t}}{b_\mathrm{mg}R_\mathrm{a}+k_\mathrm{t}^2} \left(1 - e^{-\frac{b_\mathrm{mg}R_\mathrm{a}+k_\mathrm{t}^2}{J_\mathrm{mg}R_\mathrm{a}} (t_\mathrm{s}-\Delta)} \right), \\
	\Gamma_1 &= \frac{k_\mathrm{t}}{b_\mathrm{mg}R_\mathrm{a}+k_\mathrm{t}^2} \left(e^{-\frac{b_\mathrm{mg}R_\mathrm{a}+k_\mathrm{t}^2}{J_\mathrm{mg}R_\mathrm{a}} (t_\mathrm{s}-\Delta)} - e^{-\frac{b_\mathrm{mg}R_\mathrm{a}+k_\mathrm{t}^2}{J_\mathrm{mg}R_\mathrm{a}} t_\mathrm{s}} \right).
\end{align*}
Then, similar to before, the mechanical parameters are computed as
\begin{equation*}
	b_\mathrm{mg} = \frac{1}{R_\mathrm{a}} \left( \frac{k_\mathrm{t}(1-\Phi)}{\Gamma_0 + \Gamma_1} - k_\mathrm{t}^2 \right), \quad
	J_\mathrm{mg} = - \frac{(b_\mathrm{mg} R_\mathrm{a} + k_\mathrm{t}^2)t_\mathrm{s}}{R_\mathrm{a}\text{ln}\Phi}.
\end{equation*}

Fig. \ref{system_id}(left) shows an example of data acquisition for the identification of the gearmotor parameters. 
The parameter values of the mechanical system are generally known, however to obtain a suitable plant representation, and to improve the prediction capability, certain model parameters (the mechanical Coulomb friction $\tau_\mathrm{c}$, the damping of the spring $b$) have been adjusted around their nominal values. Hence, a calibration procedure has been carried out by comparing the acquired and the simulated data and the parameter set that minimizes the root mean square error has been selected. 
Using all the identified parameters, illustrated in Table \ref{table_parameters}, model (\ref{augmented_plant_model})
has been validated on a set of independent experiments where the following quantities have been acquired via the serial communication device of the mechatronic system under analysis: the duty cycle $\delta$, the motor speed $\omega_\mathrm{m}$ and the motor current $i_\mathrm{a}$. Fig. \ref{system_id}(right) shows a sample outcome of the model validation results, obtained by comparing the identified model simulation outputs (motor current $i_\mathrm{a}$ and angular velocity $\omega_\mathrm{m}$) with the corresponding signals measured from the physical system, when the same input signal (duty cycle $\delta$) is used. Specifically, we show the responses to a staircase input. The qualitative trend of the simulated signals is close to the experimental measurements.

\section{Electrical Drive and Feedback Linearization}
\begin{figure}[t!]
\centering
\includegraphics[width=0.42\textwidth]{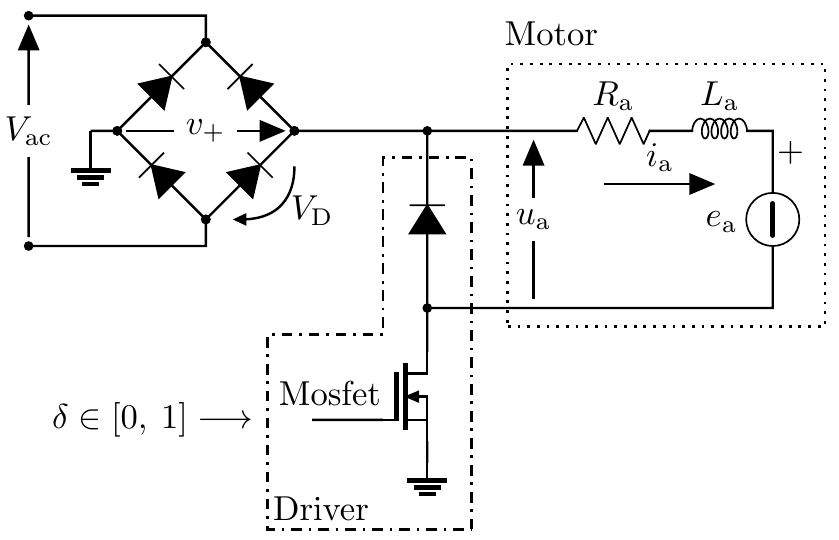}
	\caption{The electrical scheme of the driver and the equivalent motor circuit.}
	\label{Electric_scheme}
\end{figure}
\begin{figure}[t]
	\centering
	\includegraphics[width=0.45\textwidth]{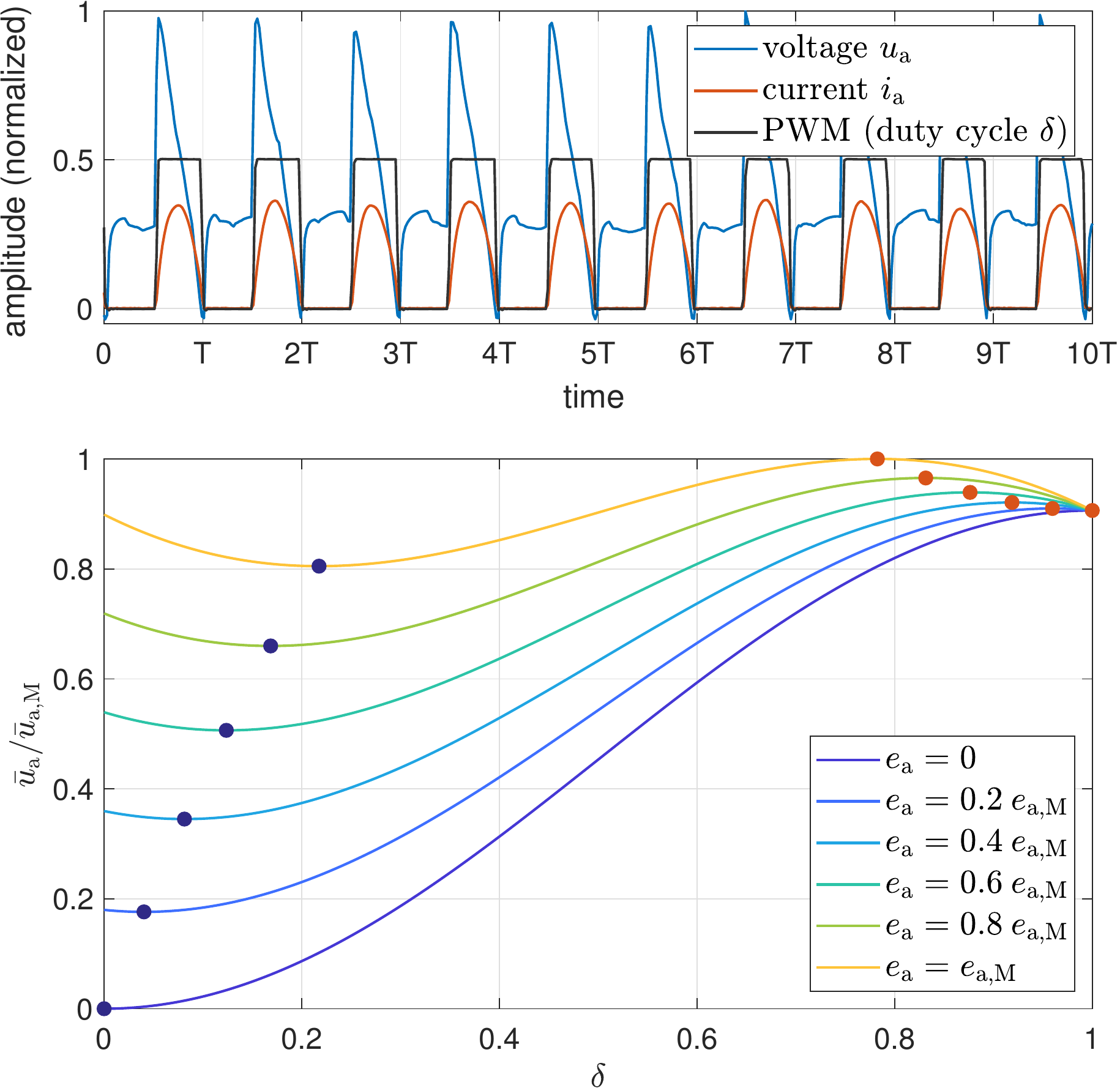}
	\caption{(Top) Time evolution of the signals related to the specific dynamics of the drive. (Bottom) Illustration of eq. (\ref{bar_u_a}) for constant BEMF values, where $e_{\rm{a,M}}$ and $\bar{u}_{\rm{a,M}}$ are normalization factors. The blue and red dots represent, respectively, the minimum and maximum points discussed in section \ref{feedback_lin_section}. }
	\label{fb_lin_fig}
\end{figure}

\subsection{Electric motor drive}
The motor driver, illustrated in Fig. \ref{Electric_scheme}, supplies the voltage to the electric motor, based on the reference signal $\delta \in \left[ 0 , \: 1 \right]$, provided by the control law. With reference to Fig. \ref{Electric_scheme}, the alternating voltage source with effective value $V_\mathrm{ac}$ is rectified by means of a Graetz bridge. The motor is controlled by chopping the non-negative semi-sinusoids $v_+$, thus modifying the average voltage depending on the on-off time of the Mosfet switching. A flyback diode, in parallel to the motor, forms a circulating path of the inductive load current.
To ease the mathematical modeling, several approximations have been carried out. The Mosfet is modelled as an ideal switch and all diodes are considered as a voltage generator when conducting current, whose voltage $V_D$ is set to the value specified in the diode data-sheet. The circuit is completed by a relay, which allows reversing the polarity of the motor when switching between opening and closing maneuvers. 
\\
The reference voltage $\delta \in \left[ 0 , \: 1 \right]$ coming from the control software governs the driver operation. Considering a single period $T$ of the rectified semi-sinusoid $v_+$, given the duty cycle $\delta \in \left[0, \: 1\right]$ during the ``Mosfet off'' portion of the duty cycle lasting $t_{\rm off} = (1-\delta) T$ seconds, the voltage $u_\mathrm{a}$ across the motor terminals is:
\begin{subequations}
	\begin{equation}
	t \in \left[0, \: (1-\delta)T \right] \Rightarrow u_\mathrm{a}(t)=
	\begin{cases}
	-V_D \:\: &\text{if } i_\mathrm{a}(t)>0 \\
	e_\mathrm{a}(t) \:\: &\text{if } i_\mathrm{a}(t)=0
	\end{cases}
	\label{eq_drive_mosfet_off}
	\end{equation}
	while during the remaining ``Mosfet on'' portion of the duty cycle lasting $t_{\rm on} = \delta T$ seconds, we have:
	\begin{equation}
	t \in \left[(1-\delta)T, \: T \right] \Rightarrow u_\mathrm{a}(t)=
	\begin{cases}
	v_+(t) \:\: &\text{if } i_\mathrm{a}(t)>0 \\
	e_\mathrm{a}(t) \:\: &\text{if } i_\mathrm{a}(t)=0. 
	\end{cases}
	\label{eq_drive_mosfet_on}
	\end{equation}
	\label{eq_driver}
\end{subequations}
An example of the corresponding signals is reported at the top of Fig. \ref{fb_lin_fig}, where it can be observed that the load is mainly resistive, since the PWM switching period is high, as compared to the electrical time constant of the armature windings. 
Therefore, to simplify eq. (\ref{eq_driver}) we can assume $i_\mathrm{a} = 0$ during the $t_{\rm off}$ phase and $i_\mathrm{a} > 0$ during the $t_{\rm on}$ phase.
Since the controller has a sampling period $T$, we aim to determine the average voltage $\bar{u}_\mathrm{a}$ of the waveform in the interval $t \in [0, \: T]$. This can be computed as
\begin{equation*}
	\bar{u}_\mathrm{a} (\delta,e_\mathrm{a}) \! = \! \frac{1}{T} \left[ \int_{0}^{(1-\delta) T} \! e_\mathrm{a} dt  +  \int_{(1-\delta) T}^{T} \sqrt{2} V_\mathrm{ac} \text{sin} \left(\frac{\pi}{T} t \right) dt \right] \! ,
\end{equation*}
where the first term is the weighted contribution of $\bar{u}_\mathrm{a}(\delta,e_\mathrm{a})$ during the $t_{\rm off}$ phase, while the weighted contribution on the $t_{\rm on}$ phase is given by the semi-sinusoid $v_+$ with amplitude $\sqrt{2} V_\mathrm{ac}$ and period $T$.
It results that $\bar{u}_\mathrm{a}$ is equal to
\begin{equation}
	\bar{u}_\mathrm{a} (\delta,e_\mathrm{a}) = e_\mathrm{a} (1-\delta) + \frac{\sqrt{2} V_\mathrm{ac}}{\pi} (1-\text{cos}(\pi \delta)).
	\label{bar_u_a}
\end{equation}
The bottom of Fig. \ref{fb_lin_fig} shows eq. (\ref{bar_u_a}) as a function of $\delta$, for $0 \le \delta \le 1$ and for constant values of $e_\mathrm{a}$, $0 \le e_\mathrm{a} \le e_{a,\rm{max}}$.

\subsection{Feedback linearization of the motor drive dynamics}
\label{feedback_lin_section}

Using a feedback linearization approach \cite{khalil}, if we manage to locally invert function (\ref{bar_u_a}), it is possible to select the duty cycle $\delta$ so that the plant seen by the controller is linear. 
Inspecting the blue and red dots in the lower plot of Fig. \ref{fb_lin_fig}, it is clear that the local inversion needs to rely on the values of $\delta$ providing the minimum and maximum of $\bar{u}_\mathrm{a}$.
Indeed, eq. (\ref{bar_u_a}) is invertible only in a range depending on $e_\mathrm{a}$. 
In particular, differentiating (\ref{bar_u_a}) with respect to $\delta$, we can determine the minimum point at $\delta=\delta_\mathrm{m}$ and the maximum one at $\delta = \delta_\mathrm{M}$ (represented by the blue and red dots of Fig. \ref{fb_lin_fig}), as a function of $e_\mathrm{a} \in [ 0,\sqrt{2}V_\mathrm{ac} )$, as follows:
\begin{subequations}
	\begin{align}
	\delta_\mathrm{m}(e_\mathrm{a}) &= \frac{1}{\pi} \text{arcsin} \left( \frac{e_\mathrm{a}}{\sqrt{2} V_\mathrm{ac}}\right) \in \left[0, \: \frac{1}{2}\right), \\
	\delta_\mathrm{M}(e_\mathrm{a}) &= 1 - \delta_\mathrm{m}(e_\mathrm{a}) \in \left(\frac{1}{2}, \: 1\right].
	\end{align}
\end{subequations}
Note that the values of $\delta_\mathrm{m}$ and $\delta_\mathrm{M}$ are well-defined since we have always $e_\mathrm{a} < \sqrt{2} V_\mathrm{ac}$. The corresponding extreme values $\bar{u}_\mathrm{a,m}(e_\mathrm{a}) = \bar{u}_\mathrm{a}(\delta_\mathrm{m}(e_\mathrm{a}),e_\mathrm{a})$ and $\bar{u}_\mathrm{a,M}(e_\mathrm{a}) = \bar{u}_\mathrm{a}(\delta_\mathrm{M}(e_\mathrm{a}),e_\mathrm{a})$ of $\bar{u}_\mathrm{a}$ can be conveniently expressed as a function of $\delta_\mathrm{m}$, omitting the dependence on $e_\mathrm{a}$ at the right-hand side, for compact notation:
\begin{subequations}
	\begin{align}
	\bar{u}_\mathrm{a,m}(e_\mathrm{a}) &= \sqrt{2}V_\mathrm{ac} \left[ (1-\delta_\mathrm{m}) \text{sin}(\pi \delta_\mathrm{m}) + \frac{1}{\pi}-\frac{\text{cos}(\pi \delta_\mathrm{m})}{\pi} \right], \\
	\bar{u}_\mathrm{a,M}(e_\mathrm{a}) &= \sqrt{2}V_\mathrm{ac} \left[ \delta_\mathrm{m} \text{sin}(\pi \delta_\mathrm{m}) + \frac{1}{\pi}+\frac{\text{cos}(\pi \delta_\mathrm{m})}{\pi} \right].
	\end{align}
	\label{u_a_m_M}
\end{subequations}
To suitably invert (\ref{bar_u_a}) based on the quantities above, define the normalized input $\tilde{\delta}$ and output $\tilde{u}_\mathrm{a}$ as follows
\begin{subequations}
\begin{align}
	\tilde{\delta}(\delta, e_\mathrm{a})&=\frac{\delta-\delta_\mathrm{m}(e_\mathrm{a})}{\delta_\mathrm{M}(e_\mathrm{a})-\delta_\mathrm{m}(e_\mathrm{a})}, \label{tilde_delta} \\
	\delta= \gamma(\tilde{\delta},e_\mathrm{a}) &= \delta_\mathrm{m}(e_\mathrm{a}) + (1-2\delta_\mathrm{m}(e_\mathrm{a}))\tilde{\delta}, \label{delta}\\
	\tilde{u}_\mathrm{a}(\tilde{\delta},e_\mathrm{a}) &= \frac{\bar{u}_\mathrm{a}(\gamma(\tilde{\delta},e_\mathrm{a}),e_\mathrm{a}) -\bar{u}_\mathrm{a,m}(e_\mathrm{a})}{\bar{u}_\mathrm{a,M}(e_\mathrm{a}) - \bar{u}_\mathrm{a,m}(e_\mathrm{a})}, 
  \label{tilde_u_a}
\end{align}
\end{subequations}
both of them taking values in $[0,1]$. 

\begin{figure}[t]
  \centering
  \includegraphics[width=0.45\textwidth]{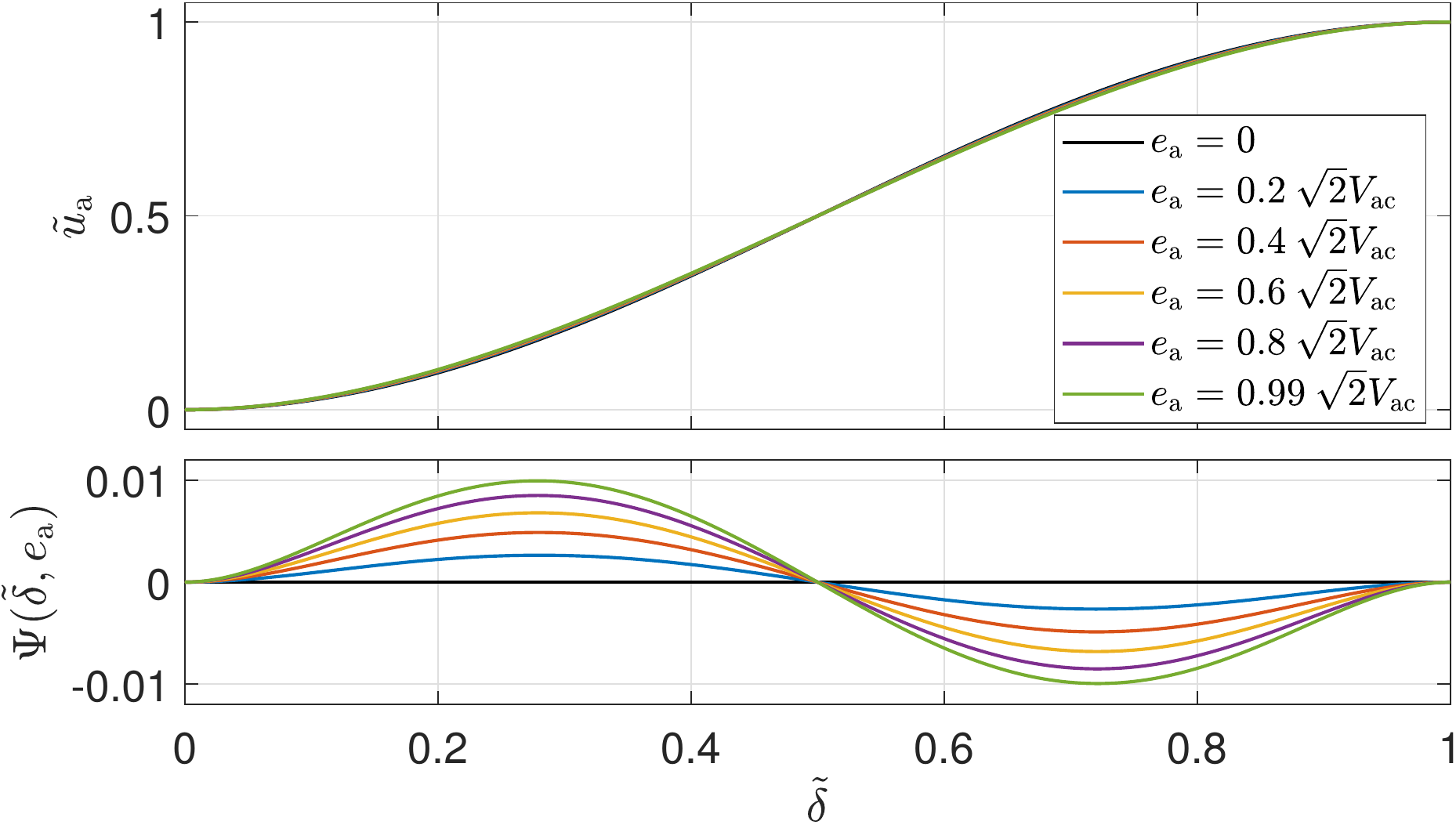}
  \caption{Function $\tilde{u}_\mathrm{a}$ in \eqref{tilde_u_a} and the mismatch function $\Psi(\tilde{\delta},e_\mathrm{a})$ defined in \eqref{approx_inverse}, represented for various values of the BEMF $e_\mathrm{a}$.}
  \label{fb_lin_fig2}
\end{figure}

In the special case $e_\mathrm{a}=0$, the expression of $\tilde u_\mathrm{a}$ in \eqref{tilde_u_a} simplifies to 
$\tilde{u}_\mathrm{a}(\tilde{\delta},0) = \frac{1 - \text{cos}(\pi \tilde{\delta})}{2} = 
\text{sin}^2\left( \frac{\pi \tilde{\delta}}{2}\right)$. More generally, we may decompose
\begin{align}
  \tilde{u}_\mathrm{a}(\tilde{\delta},e_\mathrm{a}) &= \frac{1}{2} \left( 1 - \text{cos}(\pi \tilde{\delta}) \right) + \Psi(\tilde{\delta},e_\mathrm{a}),
  \label{approx_inverse}
\end{align}
where the mismatch function $\Psi(\tilde{\delta},e_\mathrm{a})$ is small, therefore neglectable, as visible from Fig. \ref{fb_lin_fig2} and as characterized in the next lemma, whose proof is given in Section~\ref{sec:proofPSI} to avoid breaking the flow of the exposition.

\begin{lem}
\label{lem:propPSI}
For any $\tilde \delta \in [0,1]$ and any $e_\mathrm{a} \in [0,\sqrt{2}V_\mathrm{ac})$, it holds that $|\Psi (\tilde{\delta},e_\mathrm{a})| < 0.01001$.
\end{lem}

Based on the above, we can now state our main result about the inversion of function (\ref{bar_u_a}).

\begin{prop}
	\label{prop:delta}
  For any $u \in [\bar{u}_\mathrm{a,m}(e_\mathrm{a}), \: \bar{u}_\mathrm{a,M}(e_\mathrm{a})]$, selecting
  \begin{align}
  \delta =  \delta_\mathrm{m}(e_\mathrm{a}) \! + \! \frac{1\!-\!2\delta_\mathrm{m}(e_\mathrm{a})}{\pi} {\rm arccos}\left(\! 1\!-\!
  \frac{2(u -\bar{u}_\mathrm{a,m}(e_\mathrm{a}))}{\bar{u}_\mathrm{a,M}(e_\mathrm{a}) - \bar{u}_\mathrm{a,m}(e_\mathrm{a})}\!
  \right)
    \label{fb_lin_eq}
  \end{align}
  the resulting average input obtained from (\ref{bar_u_a})
   is 
   $$
   \bar{u}_\mathrm{a}(\delta, e_\mathrm{a}) = u + \psi\mbox{ with }|\psi|\le 0.01001 (\bar{u}_\mathrm{a,M}(e_\mathrm{a}) - \bar{u}_\mathrm{a,m}(e_\mathrm{a})).
   $$
   Namely, the mismatch between the requested input $u$ and the applied input $\bar{u}_\mathrm{a}$ is about one percent of the input range.
\end{prop}

\begin{proof}
  Substituing expression \eqref{fb_lin_eq} in \eqref{tilde_delta} we obtain
  $\tilde \delta(\delta, e_\mathrm{a}) = \frac{1}{\pi} {\rm arccos}\left(\! 1\!-\!
  \frac{2(u -\bar{u}_\mathrm{a,m}(e_\mathrm{a}))}{\bar{u}_\mathrm{a,M}(e_\mathrm{a}) - \bar{u}_\mathrm{a,m}(e_\mathrm{a})}\!
  \right)$. Substituting this last quantity in \eqref{approx_inverse} and using Lemma~\ref{lem:propPSI}, we obtain 
  $\tilde u_\mathrm{a} (\tilde \delta, e_\mathrm{a}) = \frac{u -\bar{u}_\mathrm{a,m}(e_\mathrm{a})}{\bar{u}_\mathrm{a,M}(e_\mathrm{a}) - \bar{u}_\mathrm{a,m}(e_\mathrm{a})} + \Psi(\tilde{\delta},e_\mathrm{a})$, with $|\Psi(\tilde{\delta},e_\mathrm{a})| < 0.01001$. The result then 
  immediately follows from \eqref{tilde_u_a}.
\end{proof}

\subsection{Proof of Lemma~\ref{lem:propPSI}}
\label{sec:proofPSI}

First, we state below a result of independent interest about a polynomial approximation of the sine function. For the proof of the result, we adopt formally certified interval arithmetic combined with formal Taylor expansion, thanks to the Coq Interval tactic \cite{martin2016proving}.

\begin{prop}
	\label{prop:sin_appr}
	For any $\alpha \in [0,1]$ the following bound holds:
	$|\sin \left(\frac{\pi}{2} \alpha \right) - \frac{1}{2} (3\alpha-\alpha^3)| < 0.02002$. Namely, the polynomial function $\frac{1}{2} (3\alpha-\alpha^3)$ approximates $\sin \left(\frac{\pi}{2} \alpha \right)$ with an error of about 2\%.
\end{prop}

\begin{proof}
	The proof is carried out using formally certified interval arithmetic software
	\cite{martin2016proving}. In particular, denoting $\xi(\alpha) := \sin \left(\frac{\pi}{2} \alpha \right) - \frac{1}{2} (3\alpha-\alpha^3)$, it is readily certified that $\xi(\alpha) \leq 0.02002$ for all $\alpha\in[0,1]$. To prove that $\xi(\alpha) \geq 0$, a certificate of positivity
	is immediate in the interval $\alpha \in [0,0.99]$, while proving non-negativity in the remaining interval $[0.99,1]$ requires proving that, in this interval, $\xi''(\alpha) \geq 0$ (so that $\xi'(\alpha)$ is non-drecreasing). Since $\xi'(1)=0$ and
	$\xi'(0.99)<0$, the above monotonicity property proves $\xi'(\alpha)\leq 0$ for
	$\alpha \in [0.99,1]$, which means the $\xi$ is therein non-increasing.
	Since $\xi(0.99)>0$ and $\xi(1)=0$, this means $\xi(\alpha)\geq 0$ for  $\alpha \in [0.99,1]$, thus completing the proof.
\end{proof}

To the end of proving Lemma~\ref{lem:propPSI}, denoting with $\sigma = \pi (1-2\delta_\mathrm{m}(e_\mathrm{a}))$, after some simplifications, we obtain from (\ref{tilde_u_a}),
\begin{equation}
	\tilde{u}_\mathrm{a}(\tilde{\delta},e_\mathrm{a}) = \frac{
	\text{tan} (\pi \delta_\mathrm{m})(\text{sin}(\sigma \tilde{\delta})-\sigma \tilde{\delta})  +1-\text{cos}(\sigma \tilde{\delta})}
	{2 - \sigma \text{tan} (\pi \delta_\mathrm{m}) }.
\label{tilde_u_a_2}
\end{equation}
Then, using $\tilde{u}_\mathrm{a}(\tilde{\delta},0)$ in \eqref{approx_inverse}, we obtain
the expression of $\Psi(\tilde{\delta},e_\mathrm{a}) = \tilde{u}_\mathrm{a}(\tilde{\delta},e_\mathrm{a}) -\tilde{u}_\mathrm{a}(\tilde{\delta},0)$ as
\begin{equation}
	\Psi \! = \! \frac{
		\text{tan} (\pi \delta_\mathrm{m}) (\text{sin}(\sigma \tilde{\delta}) \! + \! \sigma\text{sin}^2 (\frac{\pi \tilde{\delta}}{2}) \! - \! \sigma \tilde{\delta}) \! + \! \text{cos}(\pi \tilde{\delta}) \! - \! \text{cos}(\sigma \tilde{\delta})}
		{2 - \sigma \text{tan} (\pi \delta_\mathrm{m})},
	\label{mismatch_func}
\end{equation}
which clearly satisfies (see also Fig.~\ref{fb_lin_fig2}),
\begin{equation}
	\Psi(0, e_\mathrm{a}) = 0, \:\: \Psi(1, e_\mathrm{a}) = 0, \:\: \Psi(\tilde{\delta}, 0) = 0. 
  \label{psi_b}
\end{equation}

Moreover, the following symmetry is also visible from the lower representation in Fig. \ref{fb_lin_fig2}.
\begin{lem}
\label{lem1}
	For any $\alpha \in [0,1]$ and any $e_\mathrm{a} \in [0,\sqrt{2}V_\mathrm{ac})$, it holds that
$\Psi \left( \frac{1+\alpha}{2}, e_\mathrm{a} \right) = -\Psi \left( \frac{1-\alpha}{2}, e_\mathrm{a} \right)$.
\end{lem}
\begin{proof}
From eq. (\ref{mismatch_func}) consider $\text{tan} (\pi \delta_\mathrm{m}) (\sigma\text{sin}^2 (\frac{\pi \tilde{\delta}}{2}) - \sigma \tilde{\delta}) + \text{cos}(\pi \tilde{\delta})$. Substituting $\text{sin}^2(\frac{\pi \tilde{\delta}}{2})=\frac{1}{2} - \frac{1}{2} \text{cos}(\pi \tilde{\delta})$ and noting that $\text{cos}(\frac{\pi}{2} +\frac{\pi\alpha}{2})=-\text{sin}(\frac{\pi \alpha}{2})$ it follows that $\text{tan} (\pi \delta_\mathrm{m}) (\frac{1}{2} \text{sin}(\frac{\pi \alpha}{2})+\frac{\alpha}{2}) -\text{sin}(\frac{\pi\alpha}{2})= - \left[\text{tan} (\pi \delta_\mathrm{m}) (-\frac{1}{2} \text{sin}(\frac{\pi \alpha}{2}) -\frac{\alpha}{2}) +\text{sin}(\frac{\pi\alpha}{2})\right]$.
Recalling that $\sigma= \pi - 2 \pi \delta_\mathrm{m}$, the proof is completed for the remaining terms, applying the trigonometric addition formulas and noting that
$\text{tan}(\pi\delta_\mathrm{m})\text{sin}(\frac{\sigma}{2})\text{cos}(\frac{\sigma\alpha}{2})-\text{cos}(\frac{\sigma}{2})\text{cos}(\frac{\sigma\alpha}{2})=0$ since $\text{tan}(\pi\delta_\mathrm{m})\text{sin}(\frac{\sigma}{2}) = \text{cos}(\frac{\sigma}{2}) = \text{sin}(\pi\delta_\mathrm{m})$.
\end{proof}

Based on Lemma \ref{lem1}, for proving the bound in Lemma~\ref{lem:propPSI},
we may focus on its values in the range $\tilde{\delta} \in [0,\: 0.5]$, which can be parametrized by $\tilde{\delta}=\frac{1-\alpha}{2}$, $\alpha \in [0,1]$, and $e_\mathrm{a}=\sqrt{2}V_\mathrm{ac}\left(1- \frac{2}{\pi} s \right)$, $s \in (0,\frac{\pi}{2}]$. This provides
\begin{equation}
	\sup_{\substack{\tilde{\delta} \in [0,1], \\ e_\mathrm{a} \in [0,\sqrt{2}V_\mathrm{ac})}} |\Psi(\tilde{\delta},e_\mathrm{a})| = 
	\sup_{\substack{\alpha \in [0,1], \\ s \in (0,\frac{\pi}{2}]}} |\overline{\Psi}(\alpha,s)|,
	\label{sup}
\end{equation}
where $\overline{\Psi}(\alpha,s) \coloneqq \Psi\left(\frac{1-\alpha}{2},\sqrt{2}V_\mathrm{ac}\left(1- \frac{2}{\pi} s \right)\right)$.
Function $\overline \Psi$
 can be expressed as follows, after some simplifications,
\begin{align*}
	\overline{\Psi}  &= \frac{1}{2} \left( \text{sin}\left(\frac{\pi}{2}\alpha\right) - \alpha -\alpha(1-\alpha) 
	\frac{\Phi(s) - \Phi(\alpha s)}{(1-\alpha)s} \frac{1}{\Phi'(s)}  \right),
\end{align*}
where $\Phi(s) = \frac{\sin(s)}{s}$ and $\Phi'(s) = \frac{s\cos(s)-\sin(s)}{s^2}$ denotes its derivative. In particular, due to the positivity of $\Phi(s)$
and non-positivity of both $\Phi'(s)$ and $\Phi''(s)$, it holds that 
$\frac{\Phi(s) - \Phi(\alpha s)}{(1-\alpha)s} \frac{1}{\Phi'(s)}$ is non-decreasing for $s \in \left(0, \frac{\pi}{2}\right]$. Hence, due to positivity of $\alpha(1-\alpha)$, for each $\alpha \in [0,1]$, the function 
$s\mapsto \overline \Psi (\alpha,s)$ is nonincreasing
in $\left(0, \frac{\pi}{2}\right]$. Since the function is zero for $s = \frac{\pi}{2}$, then the function is everywhere positive and its supremum is given by
\begin{align*}
\overline \Psi_\mathrm{M}(\alpha) = \lim_{s \to 0^+} \overline \Psi (\alpha,s) 
= \frac{1}{2} \left( \text{sin}\left(\frac{\pi}{2}\alpha\right) - 
\frac{1}{2} (3\alpha - \alpha^3)
\right),
\end{align*}
where the right-hand side expression has been computed by applying L'H\^opital's rule three times. Finally,
applying \eqref{sup} and Proposition~\ref{prop:sin_appr}, we obtain an upper bound
on $|\Psi (\tilde \delta,e_\mathrm{a})|$ equal to $\frac{1}{2}0.02002$, 
thus completing the proof of Lemma~\ref{lem:propPSI}.

\section{Controller Design}
In this section we describe the design of a closed-loop speed controller, optimizing the performance during the opening and closing operations. The algorithm brings the barrier angular position from $\theta_0=0$ at rest to $\theta_\mathrm{f}=\pi/2$ at rest (in the opening phase) or vice-versa (in the closing phase), while satisfying a number of operating constraints. While the presented algorithm is generic, we will focus on the opening task, which is more critical due to the stringent opening time requirements.

\begin{figure}[t]
	\centering
	\includegraphics[width=0.45\textwidth]{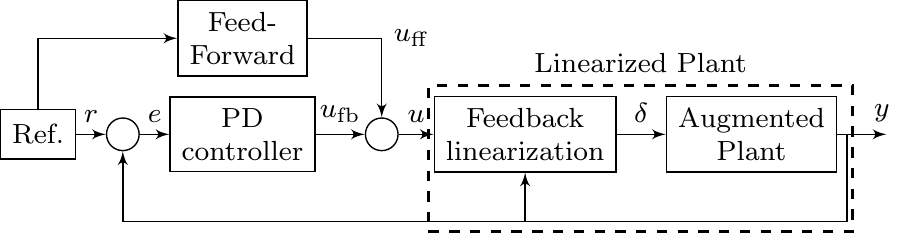}
	\caption{Blocks diagram of the control architecture.}
	\label{control_scheme}
\end{figure}

The proposed control architecture is depicted in Fig. \ref{control_scheme}. The augmented plant model \eqref{augmented_plant_model} has been defined and identified in Section \ref{Identification}.
An optimization problem is solved offline to generate a feedforward input $u_{\rm{ff}}$ and a reference trajectory $r$, specifying the desired motor angular velocity $\omega_\mathrm{m}$, which is subsequently used as reference to be tracked by the feedback controller. 
The feedback block in Fig. \ref{control_scheme} consists of a PD controller, operating at 100Hz to be compatible with the embedded software implementation. The overall control law $u$, used to compute $\delta$ from the feedback linearizing law \eqref{fb_lin_eq}, is given by
\begin{align}
\begin{split}
u(t) &= u_{\rm{ff}}(t) + u_{\rm{fb}}(t), \\ 
\end{split}
\label{eq:control_law}
\end{align}
where 
$u_{\rm{ff}}$ is an optimized feedforward input associated with the reference motion $r$ and $u_{\rm fb}$ is an error feedback stabilizer exploiting the plant measurement $y$. 
Due to Proposition \ref{prop:delta}, the dynamics from $u$ to $\bar{u}_\mathrm{a}$ is almost an identity (with a 1\% error) if $u \in [\bar{u}_\mathrm{a,m}(e_\mathrm{a}), \: \bar{u}_\mathrm{a,M}(e_\mathrm{a})]$.
The design paradigm for the feedforward and the feedback blocks of Fig. \ref{control_scheme} is explained in the next sections and can be extended to similar access automation systems.

\subsection{Reference and feedforward generation}
\label{section:Ref_generation}
The reference $r$ and the feedforward term $u_{\rm{ff}}$ are obtained by solving a constrained nonlinear optimization problem. The constraints associated to the physical limits of the system are
\begin{equation}
	\bar{u}_\mathrm{a,m}(e_\mathrm{a}) \le \bar{u}_\mathrm{a} \le \bar{u}_\mathrm{a,M}(e_\mathrm{a}), \qquad 0 \le i_\mathrm{a} \le i_\mathrm{a,M},
\end{equation}
where $\bar{u}_\mathrm{a,m}(e_\mathrm{a})$ and $\bar{u}_\mathrm{a,M}(e_\mathrm{a})$ are defined in (\ref{u_a_m_M}) and $i_\mathrm{a,M}$ is the maximum current.
To leave some input margin for the feedback action, we define 5\% tighter constraints than the actual ones. To avoid feasibility issues, the formulation with a soft constraint is introduced by adding a time-varying slack variable $\varepsilon$ and incorporating it into the cost functional. The constraints for the optimization problem become
\begin{align}
	\begin{split}
		\bar{u}_\mathrm{a,m}(e_\mathrm{a}) + \mu(e_\mathrm{a}) &\le u \le \bar{u}_\mathrm{a,M}(e_\mathrm{a}) - \mu(e_\mathrm{a}), \\
		0.05 \, i_\mathrm{a,M} - \varepsilon &\le x_1 \le 0.5 \, i_\mathrm{a,M},
	\end{split}
	\label{contraints}
\end{align}
where $\mu(e_\mathrm{a})=0.05 \, (\bar{u}_\mathrm{a,M}(e_\mathrm{a}) - \bar{u}_\mathrm{a,m}(e_\mathrm{a}))$ and $\varepsilon \in \left[0, \:\: 0.05 \, i_\mathrm{a,M} \right]$. 
Note that the motor current can only flow in one direction, which is reversed by a relay when toggling between the opening and closing phases. In the following we focus on the opening phase, the closing one being similar. Moreover, discontinuities of the input $u$ are not feasible,  
so we enforce continuity of $u$ by using as input the time derivative $v=\dot{u}$ of $u$.
The optimal control problem (OCP) is formulated as follows
\begin{subequations}
	\label{OCP}
	\begin{equation}
	\min_{\substack{x(\cdot),u(\cdot), \\ v(\cdot),\varepsilon(\cdot)}} \int_{t_0}^{t_\mathrm{f}} \!\! \| h(x(t),v(t),\varepsilon(t))\|_W^2  dt + \| h_\mathrm{f}(x(t_\mathrm{f})) \|_{W_\mathrm{f}}^2
	\label{cost_fun}
	\end{equation}
	subject to:
	\begin{align}
	&\dot{x}(t)=f(x(t),u(t)), \:\: \dot{u}(t) = v(t), \:\: \forall t \in [t_\mathrm{0},t_\mathrm{f}],
	\label{OCP1}\\
	&\psi(x(t),u(t),\varepsilon(t)) \le 0, \:\: \forall t \in [t_\mathrm{0},t_\mathrm{f}],  \label{OCP3}
	\end{align}
\end{subequations}
where $t_\mathrm{0}$ (initial time), $t_\mathrm{f}$ (terminal time) are fixed and the cost functions $h$ and $h_\mathrm{f}$ are defined as
\begin{align}
	\begin{split}
	h(x,v,\varepsilon) &= [i_\mathrm{a}, \: \theta - \theta_\mathrm{f}, \: v, \: \varepsilon]^\top, \\
	h_\mathrm{f}(x) &= [i_\mathrm{a}, \: \theta - \theta_\mathrm{f}]^\top.
	\end{split}
	\label{eq:cost_functions}
\end{align}
weighted by diagonal matrices
\begin{align}
\begin{split}
W &= \text{diag}\left(\left[ 10^{-1}, 10^{2}, 10^{-3}, 10^{7} \right] \right), \\
W_\mathrm{f} &=\text{diag}\left(\left[ 10^{-1}, 10^{2} \right] \right),
\end{split}
\end{align}
to equalize the range of the corresponding variables.
The equality constraints \eqref{OCP1} represent the dynamics of the system, where $f(x,u)$ is defined in \eqref{augmented_plant_model}, while the inequality \eqref{OCP3} comprises the constraints \eqref{contraints} with
\begin{align}
\begin{split}
\psi(x,u,\varepsilon) = \begin{bmatrix} \bar{u}_\mathrm{a,m}(e_\mathrm{a}) + \mu(e_\mathrm{a}) - u \\	
u- \bar{u}_\mathrm{a,M}(e_\mathrm{a}) +  \mu(e_\mathrm{a}) \\ 
0.05 i_\mathrm{a,M} - \varepsilon - x_1 \\
x_1 - 0.5 i_\mathrm{a,M} \\
-\varepsilon \\
\varepsilon - 0.05 i_\mathrm{a,M}
\end{bmatrix}.
\end{split}
\end{align}
The penalty related to the position error $\theta - \theta_\mathrm{f}$ in \eqref{eq:cost_functions} is needed to optimize the barrier opening time, while the one related to the current $i_\mathrm{a}$ penalizes high currents. The cost on the input $v$ ensures a sufficiently smooth control action, while the slack variable $\varepsilon$ is needed to implement the soft constraint given in \eqref{contraints}.
The OCP \eqref{OCP} is solved offline in MATLAB using the software package MATMPC \cite{matmpc}, \cite{brus1}, \cite{chen}, an open-source tool to solve nonlinear programming (NLP). In MATMPC, a NLP problem is formulated by discretizing the OCP using multiple shooting \cite{BOCK19841603} over the prediction horizon $t_\mathrm{f}$, which is divided into $N$ shooting intervals $[t_0, t_1, \dots , t_N]$.

MATMPC has been set up with a $4^{th}$ order Runge-Kutta integrator and qpOASES as QP solver \cite{ferreau2014qpoases}. The system dynamics is discretized with a sampling time $T_\mathrm{s}=0.01$s and a total number of $N=500$ shooting intervals, enabling a prediction length of $5s$. 
Given the ensuing optimal solutions $x^*(\cdot)$, $u^*(\cdot)$, we use the resulting profile $r(\cdot)=x_3^*(\cdot)$ as reference for $\omega_\mathrm{m}$ and the optimal $u^*(\cdot)$ as feedforward term, corresponding respectively to the dashed curve in Fig. \ref{experimental_results}(a) and the grey line in Fig. \ref{experimental_results}(c).

\subsection{PD feedback controller}
Based on the optimized solutions of \eqref{augmented_plant_model}, computed in \eqref{OCP} of the previous section, corresponding to trajectory $x^*$, and input $u_{ \rm ff}=u^*$, we may obtain the dynamics of the mismatch state $\tilde{x}=x^*-x$ to be stabilized by input $u_{\rm fb}$:
\begin{equation}
\begin{dcases} 
L_\mathrm{a} \dot{\tilde{x}}_1 &= - R_\mathrm{a} \tilde{x}_1 - k_\mathrm{t} \tilde{x}_3 - u_{\rm fb} \\
\dot{\tilde{x}}_2 &= \tilde{x}_3 \\
\dot{\tilde{x}}_3 &= \frac{k_\mathrm{t}}{J_\mathrm{tot}} \tilde{x}_1 - \frac{b_\mathrm{tot}}{J_\mathrm{tot}} \tilde{x}_3 +\frac{1}{J_\mathrm{tot}} w,
\end{dcases}
\label{error_model}
\end{equation}
where the exogenous signal $w$ represents the nonlinear mismatch terms comprising
\begin{itemize}
	\item the viscous friction term $b_\mathrm{tot}(x_2^*) - b_\mathrm{tot}(x_2)$;
	\item the external torque $\tau_\ell(x_2^*,x_3^*) - \tau_\ell(x_2,x_3)$.
\end{itemize}

Model \eqref{error_model} can be reduced to a lower-order system with the objective of simplifying the feedback tuning procedure.
Since the inductance $L_\mathrm{a}$ is small, we can reduce \eqref{error_model} by ignoring the (fast) electrical time constant. Fixing $L_\mathrm{a} = 0$, the first equation in (\ref{error_model}) becomes
\begin{equation}
0 = - R_\mathrm{a} \tilde{x}_1 - k_\mathrm{t} \tilde{x}_3 + u_{\rm fb} , 	\label{reduced_eq}
\end{equation}
which provides $\tilde{x}_1= \frac{u_{\rm fb} - k_\mathrm{t} \tilde{x}_3}{R_\mathrm{a}}$. This can be replaced in eq. (\ref{error_model}) to obtain
\begin{equation}
\begin{dcases}
\dot{e}_\theta \! &= e_\omega \\
\dot{e}_\omega \! &= - \left(\frac{k_\mathrm{t}^2 - b_\mathrm{tot} R_\mathrm{a}}{R_\mathrm{a} J_\mathrm{tot}} \right) e_\omega - \frac{k_\mathrm{t}}{R_\mathrm{a} J_\mathrm{tot}} u_{\rm fb} + \frac{1}{J_\mathrm{tot}} w,	
\end{dcases}
\label{reduced_model}
\end{equation} 
where $e_\theta$ and $e_\omega$ represent the angular position and velocity errors, respectively.
Considering the reduced model (\ref{reduced_model}) and defining the error dynamics $e= \left[e_\theta, \:\: e_\omega \right]\in \mathbb{R}^2$, dynamics \eqref{reduced_model} can be written as
\begin{align}
\label{eq:error_model_dynamics}
\dot{e} =& A e +B u_{\rm{fb}} + E w \\
	    :=& \begin{bmatrix} 0 &1 \\ 0 & -\frac{k_\mathrm{t}^2+b_\mathrm{tot}R_\mathrm{a}}{R_\mathrm{a} J_\mathrm{tot}} \end{bmatrix} e +\begin{bmatrix} 0 \\ -\frac{k_\mathrm{t}}{R_\mathrm{a} J_\mathrm{tot}} \end{bmatrix} u_{\rm{fb}} + \begin{bmatrix} 0 \\ \frac{1}{J_\mathrm{tot}} \end{bmatrix} w. \nonumber
\end{align}
\begin{lem}
	The pair $(A,B)$ is controllable.
	\label{lem_controllability}
\end{lem}
\begin{proof}
	This property is verified simply by noting that the controllability matrix $\mathcal{C}=\left[B| AB\right]$ has full rank.
\end{proof}

Motivated by Lemma \ref{lem_controllability}, the goal is to tune the parameters of a PD feedback control law 
\begin{align}
	u_{\rm fb} = Ke = \begin{bmatrix} k_\mathrm{p} & k_\mathrm{d} \end{bmatrix} \begin{bmatrix} e_\theta \\ e_\omega \end{bmatrix} = k_\mathrm{p} e_\theta + k_\mathrm{d} e_\omega,
	\label{eq:ufb}
\end{align}
for the closed loop system (\ref{eq:error_model_dynamics}) ensuring desirable closed-loop dynamic performance.

\subsection{PD gains tuning}
\begin{figure}
\includegraphics[width=0.18\textwidth]{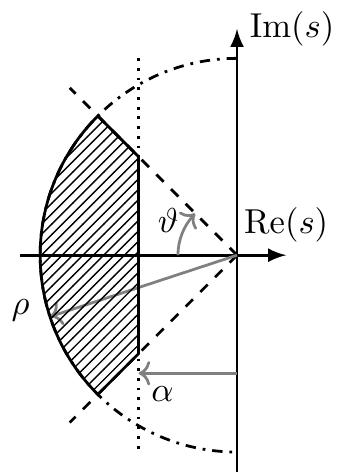}
\includegraphics[width=0.28\textwidth]{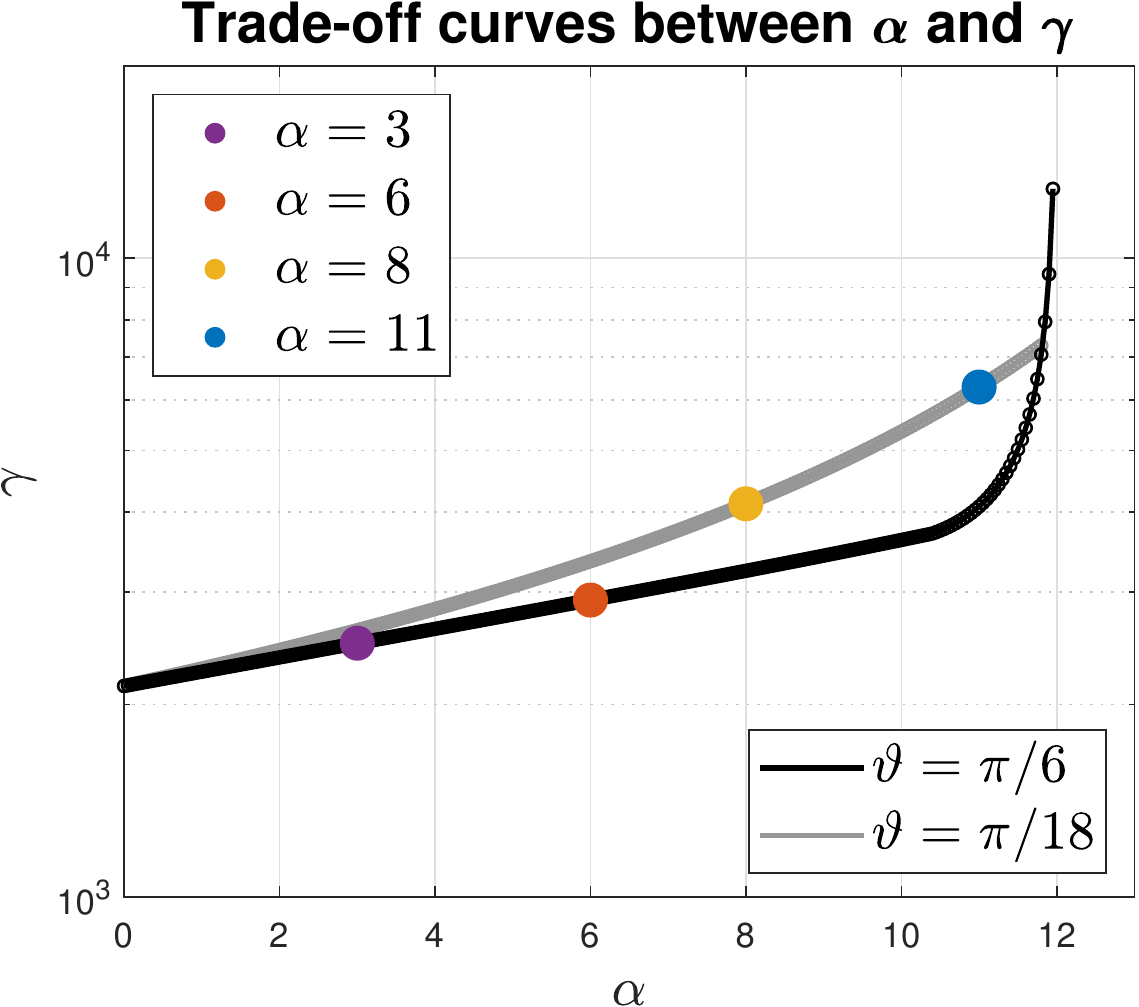}
\caption{
(Left) The shaded region where the closed-loop eigenvalues are constrained by \eqref{LMIb}-\eqref{LMId}. (Right) Trade-off curves between $\alpha$ and $\gamma$ obtained by solving the optimization problem \eqref{LMI} for model \eqref{eq:error_model_dynamics}, considering increasing values of $\alpha$ and for two different values of $\vartheta$. The colored dots correspond to the operating points chosen for the experimental results illustrated in Fig. \ref{experimental_results}.
}
\label{LMI_tuning}
\end{figure}

In this section we propose an LMI-based technique to tune the feedback controller parameters $K=\left[k_\mathrm{p}, \: k_\mathrm{d} \right]$ in \eqref{eq:ufb}. The objectives of the tuning are to guarantee the stability, to optimize the rejection of disturbance $w$, and to shape the transient performance by constraining the closed-loop eigenvalues in the shaded region of Fig. \ref{LMI_tuning} (left).

Specifically, fixing parameters $\alpha \ge 0$, $\rho > \alpha$, $\vartheta \in [0,\pi/2]$ and choosing a matrix $C$ characterizing a performance output $z=Ce$, consider the following optimization problem:
\begin{subequations}
	\label{LMI}
	\begin{align}
		&\min_{\substack{ W \in \mathbb{R}^{2 \times 2}, \\ X \in \mathbb{R}^{1 \times 2}, \\ \gamma \in \mathbb{R}}} 
		\gamma \quad \text{subject to:} \nonumber \\
		&\hspace{0.5cm} W=W^\top > 0 \label{LMIa} \\
		&\hspace{0.5cm} M + M^\top + 2\alpha W < 0 \label{LMIb} \\
		&\hspace{0.5cm} \begin{bmatrix} (M+M^\top)\text{sin}(\vartheta)  & (M-M^\top)\text{cos}(\vartheta)  \\ (M^\top-M)\text{cos}(\vartheta) & (M+M^\top)\text{sin}(\vartheta) \end{bmatrix}  \le 0 \label{LMIc} \\
		&\hspace{0.5cm} \begin{bmatrix} -\rho W  & M^\top \\ M & -\rho W \end{bmatrix} \le 0 \label{LMId} \\
		&\hspace{0.5cm} \begin{bmatrix} M+M^\top & E & WC^\top \\ E^\top &
			-\gamma I & 0 \\ CW & 0 &
			-\gamma I \end{bmatrix}  < 0   \label{LMIe},
	\end{align}
\end{subequations}
where $M:=AW+BX$, $\vartheta \in [0,\pi/2]$ and $I$ is the identity matrix of proper dimensions. 
Constraints (\ref{LMIb}), (\ref{LMIc}), (\ref{LMId}) force the closed-loop poles to lie in the shaded region of the complex plane, shown at the left of Fig. \ref{LMI_tuning}, which is the intersection of three elementary LMI regions: an $\alpha$-stability region, a disk of radius $\rho$, and a conic sector determined by $\vartheta$. The shape of this region can be adjusted using these parameters, modifying the dynamical properties of the system.

\begin{prop}
	Under Lemma \ref{lem_controllability}, for any value of $\alpha \ge 0$, $\vartheta \in [0, \pi/2]$ and $\rho>\alpha$ LMI (\ref{LMI}) is feasible. Moreover, for any feasible solution to (\ref{LMI}), selecting $K=XW^{-1}$ the following properties hold: i) the closed-loop matrix $(A+BK)$ has eigenvalues with absolute value less than $\rho$, ii) the damping factor of the poles is larger than $\text{cos}(\vartheta)$, iii) $(A+BK)$ has eigenvalues with real part smaller than $-\alpha$, iv) the $\mathcal{L}_2$ gain from $w$ to $z=Ce$ for (\ref{eq:error_model_dynamics}) with $u_{\rm fb} = K e$ is smaller than $\gamma$.
\end{prop}
\begin{proof} 
	Feasibility of (\ref{LMI}) comes from the fact that the controllability property in Lemma \ref{lem_controllability} implies a matrix $K$ that places the eigenvalues of the closed-loop system on the region of the complex semiplane defined by $\rho$, $\vartheta$ and $\alpha$. \\
	i-ii) The eigenvalues of $(A+BK)$ having an absolute value smaller than $\rho$ and the damping factor larger than $\text{cos}(\vartheta)$ are a direct application of the results in \cite[Equations (10) and (13)]{LMI2}. \\
	iii) This follows from noticing that (\ref{LMIb}) implies $(A+BK+\alpha I) W + W(A+BK+\alpha I)^\top \le 0$, which holds positive definite $W$ only if $A+BK$ has convergence abscissa smaller than $-\alpha$. \\
	iv) The proof is a standard application of the bounded real lemma and the use of quadratic Lyapunov functions. Defining $V(e)=e^\top W e$, $W=W^\top>0$ by constraint (\ref{LMIa}), performing a Schur complement on (\ref{LMIe}), left-right multiplying by $\left[ e, \: w \right]^\top$ we obtain that $\forall \left[e, \: w\right] \neq (0,0)$: 
	\begin{center}
		$ e^\top (M+M^\top) e + 2e^\top E w + \frac{1}{\gamma} z^\top z - \gamma w^\top w < 0 $.
	\end{center}
	Substituting $M=AW+BKW$ we get $\left \langle \nabla V (e), \dot{e} \right \rangle + \frac{1}{\gamma} z^\top z < \gamma w^\top w$.
	Integrating both sides, we obtain the desired bound on the $\mathcal{L}_2$ gain from $w$ to $z$, i.e. $\|z\|_2 \le \gamma \|w\|_2$ (or equivalently on the $\mathcal{H}_{\infty}$ norm).
\end{proof}

\begin{rem}
	In the presence of input saturation, the feedback system has the form considered in \cite[eqs. (1),(2)]{forni2010family}, with parameters $a_1 = 0$, $a_2 = \frac{k_\mathrm{t}^2+b_\mathrm{tot}R_\mathrm{a}}{R_\mathrm{a} J_\mathrm{tot}}$, $k=\frac{k_\mathrm{t}}{R_\mathrm{a} J_\mathrm{tot}}$ and the function $\beta(e) = \left[k_\mathrm{p}, \: k_\mathrm{d} \right] e$ 
	(using the notation of \cite{forni2010family}). Applying \cite[Thm 1]{forni2010family}, we may conclude that even with a saturated feedback $u_\mathrm{fb}$, the origin of the error system \eqref{eq:error_model_dynamics} remains globally asymptotically stable and locally exponentially stable.
\end{rem}

The LMI-based design approach \eqref{LMI} is an effective tool for performing the design of $K$. For our application, to the end of reducing as much as possible the oscillations, we select 
\begin{align}
	z=Ce=\begin{bmatrix} 0 & 1 \end{bmatrix} e = e_\omega.
	\label{eq:z}
\end{align}

The suggested use of \eqref{LMI} is to fix parameters $\rho$ and $\vartheta$ to ensure a maximum natural frequency and a minimum damping ratio. Then, for a fast decay rate, several different values of $\alpha$ can be tested to generate the corresponding trade-off curve, as reported in the right of Fig. \ref{LMI_tuning}, where we show the curves for $\vartheta=\frac{\pi}{6}$ and $\vartheta=\frac{\pi}{18}$ . The trade-off between $\gamma$ and $\alpha$ is easily seen from the resulting curves, where the solution of the optimization problem \eqref{LMI} provides the optimal gain $\gamma^*(\alpha)$ for each value of the parameter $\alpha$. 
The operating points highlighted with colored dots correspond to the feedback gains used in the experiments reported in Fig. \ref{experimental_results} of the next section.

\begin{rem}\label{rem:robust}
To certify stability of the error dynamics \eqref{eq:error_model_dynamics} in the presence of uncertain model parameters, suppose $A$ and $B$ correspond to a nominal model and that the actual matrices are not precisely known, but belong to a polytopic domain $\mathcal{D}$. Any matrix inside the domain $\mathcal{D}$ can be written as a convex combination of the vertices $A_j$ and $B_j$ of the uncertainty polytope.
We can then augment \eqref{LMI} with the following LMIs
	\begin{align}
	\label{eq:robustLMIs}
	 A_jW + B_jX  + W A_j^\top + X^\top B_j^\top < 0, \quad j = 1, \ldots p
	\end{align}
	where $p$ is the number of vertices of the polytope
	to ensure robust exponential stability of the error dynamics for any parameter in the polytope. 
\end{rem}

\section{Experimental Results}

All the experiments have been conducted on an industrial automatic barrier whose model has been identified as described in Section \ref{section:model_identification}. 
The controller has been implemented on a 8-bit microcontroller (Microchip PIC18F) mounted on a control board provided by the company developing the boom barrier. The firmware has been written in C language through MPLAB X IDE.
The reference and feedforward terms have been determined offline as described in Section \ref{section:Ref_generation}, while the PD controller and the feedback linearization have been implemented directly on the microcontroller.

\begin{figure}[t!]
	\centering
	\includegraphics[width=0.45\textwidth]{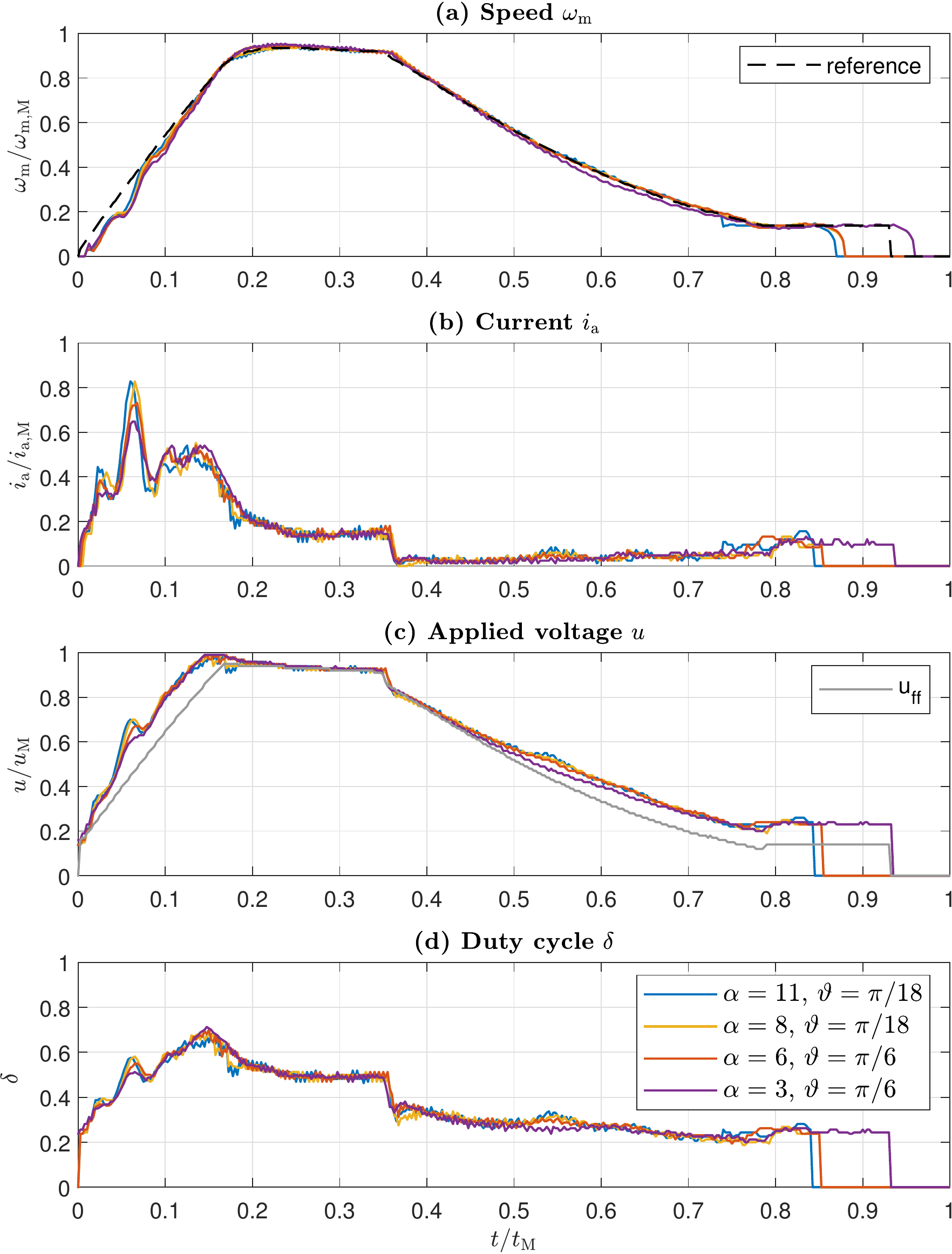}
	\caption{System responses for different values of the tuning parameters $\alpha$ and $\vartheta$. (a) reference speed $r$ and velocities $\omega_\mathrm{m}$; (b) motor currents $i_\mathrm{a}$; (c) control input $u$ delivered by the controller and (d) duty cycle $\delta$ obtained from the feedback linearization. Constants $\omega_\mathrm{m,M}$, $i_\mathrm{a,M}$, $u_\mathrm{M}$ and $t_{\rm{M}}$ are normalization factors.} 
	\label{experimental_results} 
\end{figure}

Fig. \ref{experimental_results} shows the results of the proposed control strategy and illustrates the main variables involved in the control loop. 
Following Remark~\ref{rem:robust}, rather than the nominal design in \eqref{LMI},
given the structure of $A$ and $B$ in \eqref{eq:error_model_dynamics}, we consider an uncertainty of $\pm 20\%$ on the $a_{2,2}$ element of matrix $A$ and the $b_2$ element of $B$. This provides four vertices that have been taken into account in our robust design for each one of the considered gain selections. 
The response for different values of the LMI tuning parameters $\alpha$ and $\vartheta$ show a reasonable trade-off between disturbance rejection and closed-loop performance. 
We select $\vartheta=\frac{\pi}{6}$ for lower values of $\alpha$ and $\vartheta=\frac{\pi}{18}$ for higher values, since large values of $\alpha$ may lead to optimal closed-loop gains inducing undesired oscillations, especially in the accelerating phase, probably due the mechanical backlash of the gearbox. We successfully remove the oscillations by reducing $\vartheta$ and consequently increasing the closed-loop damping ratio (see Fig. \ref{LMI_tuning}).
When $\alpha$ is too low, the tracking performance degenerates, and an undesirably large time is needed to complete the opening maneuver, due to the imperfect tracking of the reference signal.

\begin{figure}[t!]
	\centering
	\includegraphics[width=0.45\textwidth]{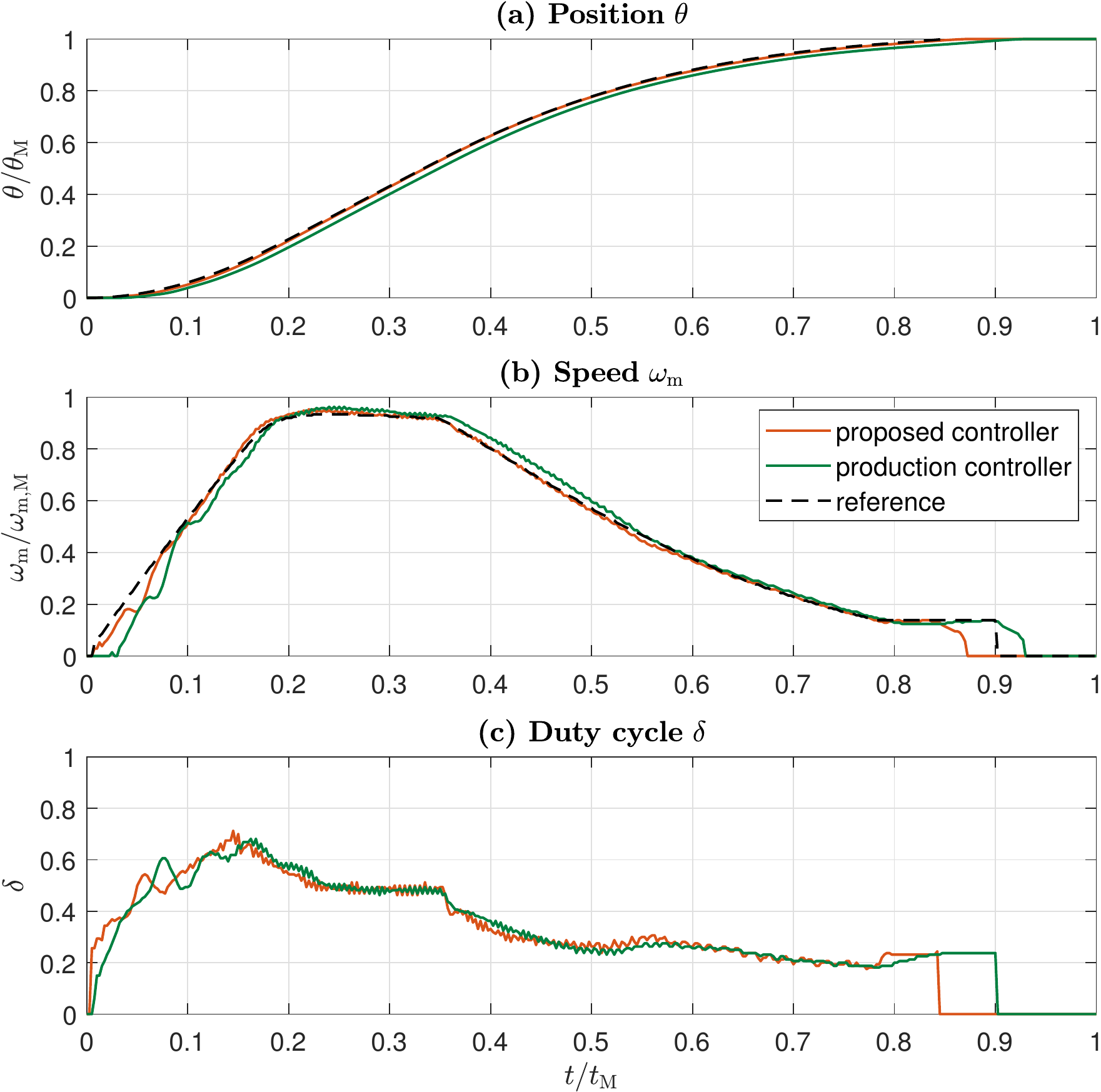}
	\caption{Output responses (a-b) and the control input responses (c) with the proposed controller and the production standard controller. Constants $\theta_\mathrm{M}$, $\omega_\mathrm{m,M}$ and $t_{\rm{M}}$ are normalization factors.}
	\label{experimental_results_comparison}
\end{figure}

Specifically, in Fig. \ref{experimental_results}(a) the dashed line shows the reference $x_3^*$ obtained by solving the optimization problem \eqref{OCP} in Section \ref{section:Ref_generation}. Fig. \ref{experimental_results}(b) shows the evolution of the current $x_1$, while Fig. \ref{experimental_results}(c) and \ref{experimental_results}(d) illustrate, respectively, the control input $u$, which is the sum of the feedforward term $u_{\rm{ff}}$ and the feedback signal $u_{\rm{fb}}$, and the duty cycle $\delta$ obtained from the feedback linearization map described in Section \ref{feedback_lin_section}.
The controller allows tracking the reference signal, compensating for disturbances and model uncertainties, resulting in a desirably small tracking error. 
Fig. \ref{experimental_results_comparison} shows a comparison between the opening maneuver responses of the proposed solution and the production standard controller.
To numerically quantify the gap between the two controllers in terms of tracking performance, we consider the NRMSE metric
\begin{equation}
	\text{NRMSE} = \frac{\lvert\lvert r_i-y_i \lvert\lvert  }{\lvert\lvert r_i - \text{mean}(r) \lvert\lvert} \quad \, i= \{1,\dots,n\} : r_i \ne 0,
\end{equation}
where $r_i=x_{3,i}^*$ and $y_i=\omega_{\rm{m},\it{i}}$, $i=1,\dots,n$ are the samples of the reference (dashed line in Fig. \ref{experimental_results_comparison}b) and the output (solid lines in the same figure). The computed NRMSE correspond to 0.1861 for the production controller and 0.0719 for the proposed showing a reduction to more than one half, 
revealing a substantial tracking accuracy improvement.
Multiple experimental tests have been carried out for a large variety of working conditions, providing excellent results, thus confirming the desirable features of the proposed solution.

\section{Conclusion}
We addressed modeling and control of a nonlinear motion system. A feedback linearization-based feedforward/feedback architecture was derived, associated with rigorous optimized performance and feasibility guarantees. This strategy is novel, it is general enough to be applicable to alternative systems sharing similar mechatronic structure, and solves the cumbersome manual gain tuning currently employed at the industrial level.

In light of the satisfactory experimental results on the considered industrial device, future work may include verifying the effectiveness of the control algorithm to explicitly account for saturation for optimizing saturated performance. We will also test the proposed strategy on alternative similar access automation systems (such as horizontal automatic gates). 
The robust results highlighted in Remark~\ref{rem:robust} will also be better investigated. Finally, a limitation of the proposed approach is the inability to account for systematic trajectory tracking errors. As this type of application is subject to repetitive maneuvers, it would be interesting to develop adaptive control techniques, despite the fact that they could be more demanding from a computational viewpoint.

\bibliographystyle{abbrv}
\bibliography{Bibliography}

\begin{thebibliography}{10}
\providecommand{\url}[1]{#1}
\csname url@samestyle\endcsname
\providecommand{\newblock}{\relax}
\providecommand{\bibinfo}[2]{#2}
\providecommand{\BIBentrySTDinterwordspacing}{\spaceskip=0pt\relax}
\providecommand{\BIBentryALTinterwordstretchfactor}{4}
\providecommand{\BIBentryALTinterwordspacing}{\spaceskip=\fontdimen2\font plus
\BIBentryALTinterwordstretchfactor\fontdimen3\font minus
  \fontdimen4\font\relax}
\providecommand{\BIBforeignlanguage}[2]{{%
\expandafter\ifx\csname l@#1\endcsname\relax
\typeout{** WARNING: IEEEtran.bst: No hyphenation pattern has been}%
\typeout{** loaded for the language `#1'. Using the pattern for}%
\typeout{** the default language instead.}%
\else
\language=\csname l@#1\endcsname
\fi
#2}}
\providecommand{\BIBdecl}{\relax}
\BIBdecl

\bibitem{Isermann}
R.~Isermann, ``Mechatronic systems - innovative products with embedded
  control,'' \emph{Control Engineering Practice}, vol.~16, no.~1, pp. 14--29,
  2008.

\bibitem{Nordin}
M.~Nordin and P.-O. Gutman, ``Controlling mechanical systems with backlash—a
  survey,'' \emph{Automatica}, vol.~38, no.~10, pp. 1633--1649, 2002.

\bibitem{Zaccarian}
M.~{Cocetti}, S.~{Donnarumma}, L.~{De Pascali}, M.~{Ragni}, F.~{Biral},
  F.~{Panizzolo}, P.~P. {Rinaldi}, A.~{Sassaro}, and L.~{Zaccarian}, ``Hybrid
  nonovershooting set-point pressure regulation for a wet clutch,''
  \emph{IEEE/ASME Transactions on Mechatronics}, vol.~25, no.~3, pp.
  1276--1287, 2020.

\bibitem{Corradini}
M.~L. Corradini and G.~Orlando, ``Robust stabilization of nonlinear uncertain
  plants with backlash or dead zone in the actuator,'' \emph{IEEE Transactions
  on Control Systems Technology}, vol.~10, no.~1, pp. 158--166, 2002.

\bibitem{lin2007robust}
F.~Lin, \emph{Robust Control Design: An Optimal Control Approach}, ser.
  RSP.\hskip 1em plus 0.5em minus 0.4em\relax Wiley, 2007.

\bibitem{hotto2007motion}
R.~Hotto, P.~D. Kahn, and L.~A. Ling, \emph{Motion control system for barrier
  drive}.\hskip 1em plus 0.5em minus 0.4em\relax US Patent 7 208 897, Apr.~24
  2007.

\bibitem{jones2008gate}
M.~W. Jones, \emph{Gate opening and closing apparatus}.\hskip 1em plus 0.5em
  minus 0.4em\relax US Patent App. 11/725 215, Sep.~25 2008.

\bibitem{levine}
W.~S. Levine, \emph{Control system applications}.\hskip 1em plus 0.5em minus
  0.4em\relax CRC press, 2018.

\bibitem{LAMBRECHTS}
P.~Lambrechts, M.~Boerlage, and M.~Steinbuch, ``Trajectory planning and
  feedforward design for electromechanical motion systems,'' \emph{Control
  Engineering Practice}, vol.~13, no.~2, pp. 145--157, 2005.

\bibitem{491410}
K.~{Ohnishi}, M.~{Shibata}, and T.~{Murakami}, ``Motion control for advanced
  mechatronics,'' \emph{IEEE/ASME Transactions on Mechatronics}, vol.~1, no.~1,
  pp. 56--67, 1996.

\bibitem{1104009}
{Kang Shin} and N.~{McKay}, ``Minimum-time control of robotic manipulators with
  geometric path constraints,'' \emph{IEEE Transactions on Automatic Control},
  vol.~30, no.~6, pp. 531--541, 1985.

\bibitem{martin2016proving}
{\'E}.~Martin-Dorel and G.~Melquiond, ``Proving tight bounds on univariate
  expressions with elementary functions in coq,'' \emph{Journal of Automated
  Reasoning}, vol.~57, no.~3, pp. 187--217, 2016.

\bibitem{Boyd}
S.~Boyd, L.~El~Ghaoui, E.~Feron, and V.~Balakrishnan, \emph{Linear matrix
  inequalities in system and control theory}.\hskip 1em plus 0.5em minus
  0.4em\relax SIAM, 1994.

\bibitem{LMI2}
M.~Chilali and P.~Gahinet, ``$\mathcal{H}_\infty$ design with pole placement
  constraints: an lmi approach,'' \emph{IEEE Transactions on Automatic
  Control}, vol.~41, pp. 358--367, 03 1996.

\bibitem{krishnan}
R.~Krishnan, \emph{Electric motor drives: modeling, analysis and
  control}.\hskip 1em plus 0.5em minus 0.4em\relax Prentice Hall, 2001.

\bibitem{leonhard2001}
W.~Leonhard, \emph{Control of electrical drives}.\hskip 1em plus 0.5em minus
  0.4em\relax Springer Science \& Business Media, 2001.

\bibitem{makkar2005}
C.~Makkar, W.~Dixon, W.~Sawyer, and G.~Hu, ``A new continuously differentiable
  friction model for control systems design,'' in \emph{Proceedings, 2005
  IEEE/ASME International Conference on Advanced Intelligent
  Mechatronics.}\hskip 1em plus 0.5em minus 0.4em\relax IEEE, 2005, pp.
  600--605.

\bibitem{ljung}
L.~Ljung, ``System identification,'' \emph{Wiley Encyclopedia of Electrical and
  Electronics Engineering}, pp. 1--19, 1999.

\bibitem{beghi1}
A.~Beghi, F.~Marcuzzi, P.~Martin, M.~Zigliotto, and F.~Tinazzi, ``Virtual
  prototyping of embedded control software in mechatronic systems: A case
  study,'' \emph{Mechatronics}, vol.~43, pp. 99--111, 05 2017.

\bibitem{khalil}
H.~K. Khalil, \emph{Nonlinear systems}.\hskip 1em plus 0.5em minus 0.4em\relax
  Prentice hall Upper Saddle River, NJ, 2002, vol.~3.

\bibitem{matmpc}
Y.~{Chen}, M.~{Bruschetta}, E.~{Picotti}, and A.~{Beghi}, ``Matmpc - a matlab
  based toolbox for real-time nonlinear model predictive control,'' in
  \emph{2019 18th European Control Conference (ECC)}, June 2019, pp.
  3365--3370.

\bibitem{brus1}
M.~{Bruschetta}, E.~{Picotti}, E.~{Mion}, Y.~{Chen}, A.~{Beghi}, and
  D.~{Minen}, ``A nonlinear model predictive control based virtual driver for
  high performance driving,'' in \emph{2019 IEEE Conference on Control
  Technology and Applications (CCTA)}, Aug 2019, pp. 9--14.

\bibitem{chen}
Y.~{Chen}, M.~{Bruschetta}, D.~{Cuccato}, and A.~{Beghi}, ``An adaptive partial
  sensitivity updating scheme for fast nonlinear model predictive control,''
  \emph{IEEE Transactions on Automatic Control}, vol.~64, no.~7, pp.
  2712--2726, July 2019.

\bibitem{BOCK19841603}
H.~Bock and K.~Plitt, ``A multiple shooting algorithm for direct solution of
  optimal control problems,'' \emph{IFAC Proceedings Volumes}, vol.~17, no.~2,
  pp. 1603--1608, 1984.

\bibitem{ferreau2014qpoases}
H.~J. Ferreau, C.~Kirches, A.~Potschka, H.~G. Bock, and M.~Diehl, ``qp{OASES}:
  A parametric active-set algorithm for quadratic programming,''
  \emph{Mathematical Programming Computation}, vol.~6, no.~4, pp. 327--363,
  2014.

\bibitem{forni2010family}
F.~Forni, S.~Galeani, and L.~Zaccarian, ``A family of global stabilizers for
  quasi-optimal control of planar linear saturated systems,'' \emph{IEEE
  Transactions on Automatic Control}, vol.~55, no.~5, pp. 1175--1180, 2010.

\end{thebibliography}

\end{document}